\documentclass[reprint,amsmath,amssymb,aps,pra,floatfix,groupedaddress,superscriptaddress]{revtex4-2}
\usepackage{graphicx} 
\providecommand{\proj}[1]{|#1\rangle\langle#1|}

\usepackage[mathscr]{euscript}
\usepackage{nicefrac}

\providecommand{\sopp}{\hat s}
\usepackage{mathtools}
\providecommand{\bL}{\boldsymbol{\Lambda}}
\providecommand{\bU}{\boldsymbol{U}}
\providecommand{\bV}{\boldsymbol{V}}

\usepackage{amsmath, amssymb, amsthm, dsfont}
\usepackage{amsfonts}

\providecommand{\ah}{\hat a}
\providecommand{\ahdag}{\hat a^\dagger}
\usepackage{float}
\usepackage[caption = false]{subfig}
\usepackage{physics}
\usepackage{wrapfig}
\usepackage[dvipsnames]{xcolor}
\definecolor{mygreen}{RGB}{44,85,17}
\definecolor{myblue}{RGB}{34,31,150}
\definecolor{myred}{RGB}{255,66,56}

\usepackage[breaklinks=true,colorlinks=true,linkcolor=myblue,urlcolor=myblue,citecolor=myblue]{hyperref}
\usepackage{svg}
\theoremstyle{definition}
\usepackage{setspace}
\graphicspath{ {./images/} }

\usepackage{times,newtx}

\begin{document}

\title{Equilibrium thermometry in the multilevel quantum Rabi model}

\author{Tabitha Doicin}
\email{tabitha.doicin@nottingham.ac.uk}
\affiliation{School of Mathematical Sciences, University of Nottingham, University Park, Nottingham NG7 2RD, UK}

\author{Luis A. Correa}
\affiliation{Instituto~Universitario~de~Estudios~Avanzados~(IUdEA),~Universidad~de~La~Laguna,~La~Laguna~38203,~Spain}
\affiliation{Secci\'on de F\'isica, Facultad de Ciencias, Universidad de La Laguna, La Laguna 38203, Spain}

\author{Jonas Glatthard}
\affiliation{School of Physics and Astronomy, University of Nottingham, Nottingham NG7 2RD, United Kingdom}

\author{Andrew D. Armour}
\affiliation{School of Physics and Astronomy, University of Nottingham, Nottingham NG7 2RD, United Kingdom}
\affiliation{Centre for the Mathematics and Theoretical Physics of Quantum Non-Equilibrium Systems, University of Nottingham, Nottingham NG7 2RD, UK}%

\author{Gerardo Adesso}
\affiliation{School of Mathematical Sciences, University of Nottingham, University Park, Nottingham NG7 2RD, UK}
\affiliation{Centre for the Mathematics and Theoretical Physics of Quantum Non-Equilibrium Systems, University of Nottingham, Nottingham NG7 2RD, UK}%

\date{February 2026}

\begin{abstract}
The temperature sensitivity of a probe in equilibrium can be gauged by its thermal quantum Fisher information (QFI).
It is known that probes exhibiting degeneracy in their energy-level structure can achieve larger sensitivities, while probes with a more uniform spectrum may remain sensitive over a broader temperature range. 
Here, we study the thermometric performance of a multilevel quantum Rabi model in which two well-separated atomic manifolds of near-degenerate levels couple to a single cavity mode.
We generalise the standard quantum Rabi treatment in the adiabatic regime to find an approximate closed-form expression for the thermal QFI. We then characterise two complementary limits. On the one hand, a large dark-state manifold (dark-manifold saturation) produces a robust peak in thermal sensitivity due to bright--dark population transfer. Such increase in sensitivity is further maximised at an intermediate light--matter coupling strength.
Maximising instead the number of bright states (bright-manifold saturation) generates a broadband thermal response that becomes increasingly stable under random light--matter couplings as the number of levels is increased. The rich spectral structure of our cavity-QED model thus makes it a versatile and sensitive equilibrium thermometer over a broad range of temperatures.
\end{abstract}

\maketitle

\section{Introduction}\label{sec:intro}

Quantum thermometry \cite{Therm_Review,DePasquale2018,campbell2025roadmap} aims to estimate the temperature of an equilibrium sample by coupling it to a quantum probe. If the probe is in a canonical equilibrium state, its thermometric precision is fundamentally limited by its quantum Fisher information (QFI) \cite{QFI_original_1,QFI_original_2,PhysRevLett.72.3439,QFI_review,quantumestimation} and, in turn, this precision is set by equilibrium energy fluctuations \cite{PhysRevE.83.011109,Therm_3_optimal}. Hence, the energy-level structure of the probe may be engineered to provide substantial precision gains \cite{Therm_3_optimal,campbell2018precision,PhysRevA.97.063619,mok2021optimal,abiuso2024optimal,PhysRevA.111.052216,mukherjee2019enhanced,glatthard2022bending}. In the provably `ideal' \cite{reeb2015tight,Therm_3_optimal} case of a two-level probe with maximally degenerate excited manifold, the QFI attains its largest peak at an operating temperature set by the level spacing and degeneracy, but its thermal response is correspondingly narrow \cite{Therm_3_optimal,campbell2018precision,PhysRevA.97.063619,PhysRevA.111.052216}. More balanced degeneracy allocations typically yield a broader response at the expense of reduced peak sensitivity \cite{mok2021optimal,abiuso2024optimal}. Hence, we can either increase peak sensitivity through strong degeneracy imbalance \cite{ullah3}, or broaden the useful temperature range by supporting many thermally active transitions across a spread of energies \cite{ullah1,ullah2}.

\begin{figure*}[t]
\centering
\includegraphics[width=0.8\linewidth]{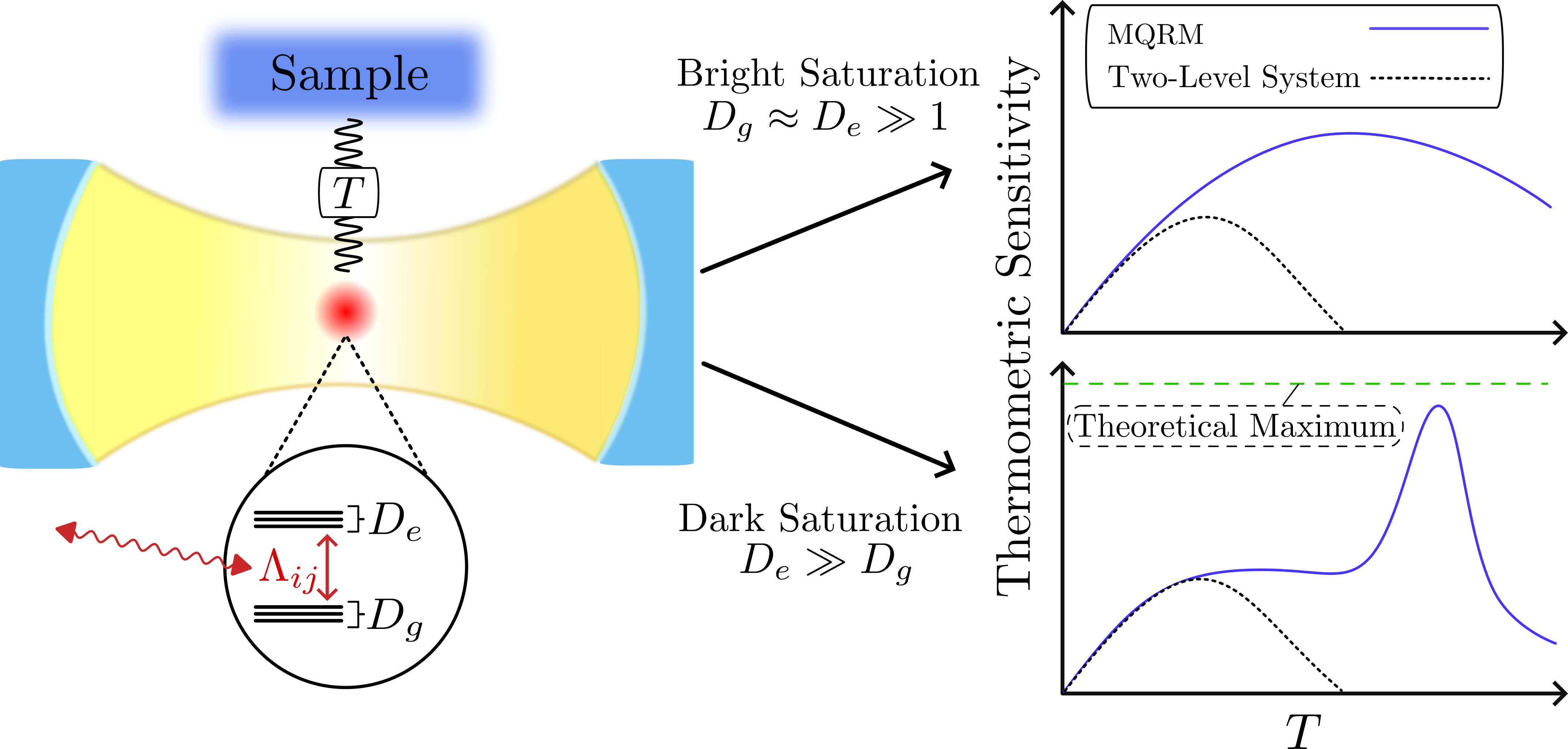}
\caption{\textbf{Schematic overview of large--manifold MQRM thermometry} and the two limiting cases we consider. A multilevel atom with near-degenerate ground and excited manifolds of sizes $D_g$ and $D_e$ couples to a single cavity mode through a general complex light--matter coupling matrix $\Lambda_{ij}$, and is brought into thermal equilibrium with a sample at temperature $T$. The resulting thermal sensitivity depends strongly on how the spectrum is partitioned into bright and dark sectors. \textbf{Bright-manifold saturation} ($D_g\approx D_e\gg1$) distributes sensitivity across many bright transitions, producing a broad response. \textbf{Dark-manifold saturation} ($D_e\gg D_g$) generates a dominant peak associated with bright--dark population transfer, that can approach the theoretical maximum for a matched effective degeneracy.}
\label{fig:schematicdiagram}
\end{figure*}

Here, we explore these two complementary situations in the context of a multilevel generalisation of the much celebrated quantum Rabi model (QRM) \cite{Jaynes1963,Xie2017,Braak2017}, where two nearly degenerate atomic manifolds couple to a single bosonic mode through a general complex coupling matrix \cite{TommasoQRM,MQRM}. A multilevel quantum Rabi model (MQRM) of this type provides a minimal cavity-QED setting in which collective bright superpositions hybridise strongly with the field, while dark states remain weakly affected, yielding a rich structured spectrum with controllable degeneracies. Related multilevel light--matter spectra appear in several ultrastrong coupling platforms \cite{Review_USC_Polariton}, including cavity embedded two-dimensional electron gases supporting inter-Landau-level polaritons \cite{USC_Polariton,GaAsQW,InAsQW,HeliumQW}, as well as semiconductor quantum dots with large exciton degeneracies  \cite{PbSorSeQDpaperoriginal,TommasoQRMExperimental,64degenQD,64DegenQD2}. While we are not aiming to model these systems microscopically, our focus will be on investigating how the bright--dark manifold structure and generic multichannel couplings can shape general equilibrium thermometric performance.

A central difficulty is that direct numerical thermometry in large MQRM manifolds is computationally very expensive, as the cavity field renders the spectrum infinite, and the atomic manifold can introduce a large number of coupled transitions. We therefore work within the adiabatic regime, where the atomic frequency is small relative to that of the cavity \cite{RegimesQRM_1,Xie2017,Braak2017}. In this limit we derive a multilevel adiabatic approximation (AA) that generalises the standard QRM treatment \cite{AA_1,AA_2,AA_3}. We first bring the light--matter coupling into a `superradiant' basis, which isolates bright QRM-like doublets with effective couplings and separates excess states into a dark sector \cite{MQRM} and then diagonalise the model under the assumption of total atomic degeneracy. Finally, we reintroduce atomic structure and small intraband atomic detunings perturbatively, 
yielding a tractable and effective approximate diagonalisation.

Within this AA, we obtain an explicit expression for the thermal QFI in terms of bright and dark spectral contributions. This form naturally separates thermometric features associated with (i) intra-doublet bright splittings, (ii) inter-doublet bright--bright processes, and (iii) bright--dark population transfer, allowing distinct parts of the QFI curve to be traced back to specific classes of transitions. To evaluate the QFI efficiently at large manifold size, we introduce an energy-ordered AA truncation scheme, where rather than truncating to fixed photon number, we retain all terms relating to levels up to a controlled energy cutoff. This provides fast convergence over the temperature range of interest and makes large ensembles of such disordered systems computationally feasible to study. 

We analyse two complementary large-manifold limits (see concept diagram in Fig.~\ref{fig:schematicdiagram}) performing Monte Carlo sampling of realisations of the light--matter couplings and intraband broadening to assess ensemble stability. This allows us to resolve both the typical QFI profile and the extent of sample-to-sample fluctuations under disorder. At dark-manifold saturation conditions a small bright sector couples to a large, nearly degenerate dark band, producing a dominant high-temperature QFI peak driven by the large degeneracy imbalance. As the dark degeneracy grows at fixed bright structure, the sensitivity peak approaches the ideal-thermometer benchmark at the matched degeneracy and effective gap. Given the existence of a small bright manifold, we also identify an intermediate coupling window in which spectral reshuffling isolates the dominant bright--dark gap and yields a higher peak QFI ratio to the matched ideal benchmark than in the weak coupling limit. In the complementary bright-manifold saturation regime the number of bright doublets is maximised, and sensitivity is distributed across a dense set of inter-doublet transition energies determined by the singular value spectrum. This yields a broadband QFI response that becomes progressively self-averaging with increasing manifold size, although its peak sensitivity typically remains below that of the dark-saturated configuration. 

Finally, we also examine thermometric sensitivity when experimental access is restricted to either the atomic or the cavity degrees of freedom. 
This is practically relevant, as the global QFI assumes access to measurements on the full joint light--matter state and, therefore, gives an upper bound rather than a directly accessible figure of merit. Importantly, we find that the atomic reduced state typically retains most of the thermometrically relevant global response, which indicates that the main enhancement can remain visible even without full access to the joint light--matter state.

This paper is organised as follows. Sec.~\ref{sec:system} introduces the MQRM and derives the multilevel adiabatic approximation. Sec.~\ref{sec:metro} gives the AA expression for the thermal QFI, and Sec.~\ref{sec:discussion} uses it to analyse bright- and dark-saturation thermometry, respectively, including stability under disorder, typical behaviour in large random ensembles, and reduced-state thermometry. Finally, in Sec.~\ref{sec:conclusion}, we provide our conclusions and outlook.

\section{The System}\label{sec:system}

\subsection{Model Hamiltonian}

We consider a multilevel quantum Rabi model (MQRM) \cite{MQRM} where the traditional two-level atom \cite{Jaynes1963,Braak2017} is replaced by two bands of nearly degenerate states, with Hamiltonian given by ($\hbar=1$)
\begin{equation}    \label{MQRMhamiltonian}
\begin{aligned}
    \hat H&=\omega_f \hat a^\dagger \hat a+ \omega_a\sum_{j=1}^{D_e}\proj{e_j} \\ &+\varepsilon\sum_{j=1}^{D_e}\delta^e_{j}\proj{e_j}+\varepsilon\sum_{i=1}^{D_g}\delta^g_{i}\proj{g_i} \\
   &+\sum_{i=1}^{D_g}\sum_{j=1}^{D_e}\left(\Lambda_{ij}\sopp_{ij}+\Lambda_{ij}^*\sopp_{ij}^\dagger\right)\left(\hat a+\hat a^\dagger\right).
   \end{aligned}
\end{equation}
The model is comprised of a single-field mode with annihilation operator $\ah$ and bare frequency $\omega_f$, coupled to a multilevel atom with $D_e$ excited states $\{\ket{e_1},\cdots,\ket{e_{D_e}}\}$, with bare energies $\{\omega_a +\varepsilon\delta^e_{1},\cdots,\omega_a +\varepsilon\delta^e_{{D_e}}\}$, and $D_g$ ground states $\{\ket{g_1},\cdots,\ket{g_{D_g}}\}$, with bare energies $\{\varepsilon\delta^g_{1},\cdots,\varepsilon\delta^g_{D_g}\}$. The atomic transition operators within the light--matter coupling term are given as 
\begin{equation}
    \sopp_{ji}\coloneqq\ket{g_j}\bra{\vphantom{g_j}e_i}.
\end{equation}
We assume that the detunings are constrained to  $\delta^e_{i},\delta^g_{i}\in[-1,1]$, so that $\varepsilon$ sets the scale of the overall spread of the excited and ground states about $\omega_a$ and $0$, respectively. In the cases we will consider, the model is assumed to be either degenerate ($\varepsilon=0$) or nearly so, with $\varepsilon/\omega_a\ll 1$. A complex light--matter coupling matrix $\bold \Lambda$, with elements $\Lambda_{ij}$, specifies how each ground-to-excited atomic transition couples to the field quadrature $\ah+\ahdag$. 

We can switch to a more convenient `superradiant' basis for probing the system by performing a singular value decomposition (SVD) on the coupling matrix $\boldsymbol\Lambda$, such that
\begin{equation}\label{eq:svd}
\boldsymbol{\Lambda}=\bU\boldsymbol{\lambda}\boldsymbol{V}^\dagger
\end{equation}
for unitary $\bU\in \mathbb{C}^{D_g\times D_g}$ and $ \bV\in \mathbb{C}^{D_e\times D_e}$.
Let us define $M = \text{rank}(\bL)\leq \min(D_g,D_e)$ and $N = \max(D_g,D_e)$, so that the interaction term of the Hamiltonian in SVD form in the atomic subspace becomes 
\begin{eqnarray}
    \hat H_\text{int} = \sum_{k=1}^{M}\lambda_k (\ket{G_k}\bra{E_k} + \ket{E_k}\bra{G_k})(\ah+\ahdag),
\end{eqnarray}
with the new atomic states defined as 
\begin{eqnarray}
    \ket{G_k} = \sum_{i=1}^{D_g} U_{ik} \ket{g_i \vphantom{e_j}}  
 \ \mathrm{and} \ \ket{E_k} = \sum_{j=1}^{D_e} V_{jk} \ket{e_j}.
\end{eqnarray}

As we can see, in this superradiant basis the light--matter interaction between the $k$-th doublet pair $\ket{G_k},\ket{E_k}$ and the cavity is mediated by the singular values $\lambda_k$, with $\lambda_1$ being the largest and decreasing for $k>1$. We say the SVD basis state pairs $\{\ket{G_1},\ket{E_1}\}, \cdots \{\ket{G_M},\ket{E_M}\}$ are `bright', due to their enhanced interaction to the cavity mode. In contrast, states $\ket{G/E}_k$ for $M<k\leq N$ have no light--matter coupling at all since they do not show up in the interaction term, and are thus said to be completely `dark'.

The MQRM Hamiltonian Eq.~\eqref{MQRMhamiltonian} can, therefore, be cast into the much more convenient form
\begin{equation}\label{eq:full_hamiltonian}
\begin{aligned}
    \hat H&=\omega_f \hat a^\dagger \hat a + \sum_{k=1}^{M}\lambda_k (\ket{G_k}\bra{E_k} + \ket{E_k}\bra{G_k})(\ah+\ahdag)\\ 
    &+ \omega_a\sum_{j=1}^{D_e} \proj{E_j} +\varepsilon\sum_{j,k,l=1}^{D_e}\delta^e_{j}V^*_{jk}V_{jl}\ket{E_k}\bra{E_l} \\
    &+\varepsilon\sum_{i,k,l=1}^{D_g}\delta^g_{i}U^*_{ik}U_{il}\ket{G_k}\bra{G_l}.
\end{aligned}
\end{equation}
Unfortunately, this basis change comes at the cost of complicating the previously diagonal intraband detuning terms, which now serve as the only means to couple the different superradiant states, including the dark manifold, and induce avoided crossings of order $\sim \varepsilon$ \cite{MQRM}. However, as we previously assumed that the levels within each band are nearly degenerate, i.e., `the narrow bandwidth approximation', these effects are minimal and can be managed.

\subsection{Adiabatic approximation} 
\label{AA_section}
We now analyse the MQRM in the adiabatic regime \cite{Braak2017,RegimesQRM_1,Xie2017}, where $\omega_a/\omega_f\ll1$, such that the atomic frequency is much smaller than that of the oscillator. Since our model is more complex than the conventional quantum Rabi model (QRM) \cite{AA_1,AA_2,AA_3}, deriving adiabatic-limit energies requires care and some extra approximations. 
To begin with, we consider the zeroth-order AA \cite{AA_1,RegimesQRM_1}, where inter and intra band splittings are set to zero ($\omega_a=\varepsilon=0$) and we only keep the cavity energy and light--matter interaction terms. In the superradiant basis the reduced Hamiltonian reads
\begin{equation}
    \hat H_\text{red} = \omega_f \hat a^\dagger \hat a + \sum_{k=1}^{M}\lambda_k (\ket{G_k}\bra{E_k} + \ket{E_k}\bra{G_k})(\ah+\ahdag).
\end{equation}
Let us define the `bright coupling' operator $\hat{Q}$ as
\begin{equation}
    \hat{Q} = \sum_{k=1}^M \lambda_k (\ket{G_k}\bra{E_k} + \ket{E_k}\bra{G_k}),
\end{equation}
with eigenstates
\begin{equation}
    \hat{Q}\ket{\lambda_k^\pm} = \pm \lambda_k \ket{\lambda_k^\pm},  \quad \ket{\lambda_k^\pm} = \frac{1}{\sqrt{2}}(\ket{E_k} \pm \ket{G_k}),
\end{equation}
for $k\in[1,M]$.
Let us further make an \textit{ansatz} separating atomic and cavity sectors as $\ket{\upsilon} = \ket{\lambda_k^{\pm}}\otimes\ket{\phi}$. Factoring out the atomic parts of the Hamiltonian, we are left with the displaced oscillator
\begin{equation}
    \left(\omega_f\ahdag \ah \pm \lambda_k (\ahdag + \ah) \right) \ket{\phi} = E\ket{\phi}.
\end{equation}
Completing the square yields a displaced oscillator eigen-decomposition on the cavity sector \cite{AA_1}, with eigenstate pairs satisfying
\begin{equation}
    \ket{\upsilon_{k\pm}^n} = \ket{\lambda_k^\pm}\otimes\ket{n_k^{\pm}}, \quad \ket{n_k^\pm} = \hat{D}\left(\mp \lambda_k/\omega_f\right)\ket{n},
\end{equation}
where $\hat{D}(\alpha) = e^{\alpha\ahdag - \alpha^* \ah}$ is the displacement operator, and the states have twofold-degenerate energy levels
\begin{equation}
    E_k(n) = \omega_f \big(n-\lambda_k^2/\omega_f^2\big).
    \label{0thorderAAbright}
\end{equation}
Compared to the QRM, this model yields a ladder of bright twofold-degenerate adiabatic levels within the zeroth order AA, with energies depending on their own specific light--matter coupling. The dark sector, with singular values of $\lambda_k=0$, remains uncoupled and has free cavity $N-M$ degenerate energies $E(n) = \omega_fn$  for $D_g>D_e$ (or $E(n) = \omega_f\,n+\omega_a$ for $D_g<D_e$).

Before we take the next step and write out the entire Hamiltonian within this $0$-th order AA displaced oscillator basis, it is convenient to consider the `SVD detuning' approximation. 
Let us treat the small detuning terms in Eq.~\eqref{eq:full_hamiltonian} for the excited band, i.e.,
\begin{equation}
\hat{\Theta}_E := \varepsilon\sum_{j,k,l=1}^{D_e}\delta^e_{j}V^*_{jk}V_{jl}\ket{E_k}\bra{E_l},
\end{equation}
as a perturbation. Assuming $D_e>M$ (so that the band dimension exceeds the population of the bright manifold) and restricting first to the bright subspace, where $k\leq M$, we know from the $0$-th order AA that the $k$-th bright eigenstate will be made up of a superposition of the $k$-th bright doublet pair elements $\ket{G_k},\ket{E_k}$, to zeroth order in $\omega_a$. To first order in perturbation theory, we retain
\begin{equation}
    \hat{\Theta}_E \approx \sum_{k=1}^{M}\Delta^e_k \ket{E_k}\bra{E_k} , \quad \Delta^e_k = \varepsilon\sum_{j=1}^{D_e}\delta_j^e|V_{jk}|^2,
\end{equation}
with $\Delta_k^{e}$ absorbing $\varepsilon$, giving the `SVD averaged' detuning contribution to the energies in the excited manifold. The same procedure can be followed for the ground-state intraband detuning term. This approximation works best when far from any crossings between any two energy levels, i.e., $\Delta E \gg\varepsilon$. We will adopt it globally, however, accepting that it is least accurate at avoided crossings. This is due to the associated error being bounded at order $\varepsilon$ and remaining negligible throughout the parameter ranges we consider ($\varepsilon/\omega_a\ll1$). Within the dark sector, the unperturbed manifold is quasi-degenerate, so all $\hat\Theta_{E/G}$ elements contribute on the same scale $\mathcal{O}(\varepsilon)$, however, as full numerics confirm (Fig.~ \ref{fig:AAverificationQFI}), thermometric precision is insensitive to this fine structure, so we simply ignore it. 

The dark sector is now fully decoupled from the bright one and is diagonal. We denote restrictions to the dark and bright subspaces by subscripts $D$ and $B$, respectively. The approximate dark-sector Hamiltonian is therefore
\begin{eqnarray}\label{eq:dark-Hamiltonian}
    \hat H_{D} \simeq \omega_f \ahdag \ah + \frac{\omega_a}{2}\sum_{k=1}^{N-M}(1+p)\ket{D_k}\bra{D_k},
\end{eqnarray}
where we define $p=1$ for if the excess states are part of the excited band ($D_e>D_g$), or $p=-1$ if the excess states are part of the ground band ($D_g>D_e$). For clarity, the excess dark states have been relabeled as $\ket{G_k/E_k}\rightarrow\ket{D_k}$ for $k>M$. The dark block is diagonal at this order within the AA and so the energies are given by
\begin{subequations}
\begin{align}
    \Lambda_{k\emptyset}^{n} &= n\,\omega_f + \frac{\omega_a}{2}\,(1+p),\label{AAdarkenergies}\\
    \ket{\Lambda_{k\emptyset}^n} &= \ket{D_k}\otimes\ket{n} \qquad k\in[M+1,N].
\end{align}
\end{subequations}

For the bright sector, we project the remaining splitting terms onto the bright block, which we deonte by $(\cdots)_B$. 
Within the SVD detuning approximation this gives
\begin{align}
\left(\hat H-\hat H_\text{red}\right)_B
&\approx \sum_{k=1}^{M}
\left[
(\omega_a+\Delta^e_k)\ket{E_k}\bra{E_k}
+\Delta^g_k\ket{G_k}\bra{G_k}
\right] \nonumber \\
&=
\frac{1}{2}\sum_{k=1}^{M}
\Big[
(\omega_a+\Delta_k^+)
\left(
\ket{\lambda_k^+}\bra{\lambda_k^+}
+\ket{\lambda_k^-}\bra{\lambda_k^-}
\right) \nonumber \\
& \quad + (\omega_a+\Delta_k^-) \left( \ket{\lambda_k^+}\bra{\lambda_k^-} +\ket{\lambda_k^-}\bra{\lambda_k^+}
\right) \Big],
\end{align}
with the detunings recast as
\begin{equation}
\Delta_k^{\pm} = \Delta_k^e \pm \Delta_k^g.
\end{equation}
This part of the Hamiltonian is written in the atomic sector and only couples terms belonging to the same superradiant doublet. However, it does contain many contributions that couple different photon-numbered blocks, due to the overlap between different displaced oscillator eigenstates. The adiabatic self-energy approximation can be now used to eliminate such couplings, as done in the standard 
QRM AA derivation \cite{AA_1,AA_2,AA_3}. Namely, we assume that $\omega_a/\omega_f\ll1$, and only retain terms diagonal in photon number $n$ and eliminate $n\neq m$ couplings. After truncation, the remaining Hamiltonian for the bright sector is now set in a block-diagonal form \cite{AA_1}, and we project it onto the subspace spanned by $\{\ket{\upsilon_{k+}^n},\ket{\upsilon_{k-}^n}\}$, with 
\begin{eqnarray}
    \hat{H}^\text{eff}_{n,k} = \begin{pmatrix}
    E_k(n) + \frac{1}{2}(\omega_a +\Delta_k^+) & \frac{1}{2}(\omega_a+\Delta_k^-)\bra{n_k^-}\ket{n_k^+}\\
    \frac{1}{2}(\omega_a+\Delta_k^-)\bra{n_k^-}\ket{n_k^+} & E_k(n) + \frac{1}{2}(\omega_a +\Delta_k^+)
\end{pmatrix} .
\end{eqnarray}
The displaced oscillator overlaps are given by \cite{deoliveirafockstates} 
\begin{eqnarray}
    \braket{n_k^-}{n_k^+}= e^{-2\lambda_k^2/\omega_f^2}\,\mathcal{L}_n\left(4\lambda_k^2/\omega_f^2\right),
\end{eqnarray}
where $\mathcal{L}_n$ denotes the $n$-th Laguerre polynomial. Using the above, we can calculate the eigenenergies as
\begin{equation}
\begin{aligned}
    \Lambda_{k\pm}^n &= \omega_f\left(n-\frac{\lambda_k^2}{\omega_f^2}\right) + \frac{1}{2}(\omega_a+\Delta_k^+) \\ 
    &\pm \frac{1}{2}(\omega_a+\Delta_k^-)\,e^{-2\lambda_k^2/\omega_f^2}\,\mathcal{L}_n\left(4\lambda_k^2/\omega_f^2\right),
\end{aligned}
    \label{AAbrightenergies}
\end{equation}
for $k\in[1,M]$, with eigenstates given by 
\begin{eqnarray}
    \ket{\Lambda^n_{k\pm}} = \frac{1}{\sqrt{2}}\left(\ket{\upsilon_{k+}^n}\pm\ket{\upsilon_{k-}^n}\right).
\end{eqnarray}

We have thus obtained displaced oscillator ladders for each bright superposition in the fully degenerate limit, and then introduced the atomic splittings as perturbative corrections to this dressed basis. 
This procedure yields an effective Hamiltonian in which the dark block is effectively decoupled and diagonal, and the bright doublets can be treated independently, while still retaining information about the underlying detuning profile through a set of effective shifts.
We can verify that this spectrum does reduce to that of the QRM with the correct parameters \cite{AA_1,RegimesQRM_1} ($M\rightarrow1,\Delta_1^{\pm}\rightarrow0,\lambda_1=g$), with energies
\begin{equation}
    E^n_{\pm} = \omega_f\left(n-\frac{g^2}{\omega_f^2}\right) + \frac{1}{2}\omega_a \pm \frac{1}{2}\omega_a\,e^{-2g^2/\omega_f^2}\,\mathcal{L}_n\left(4\,g^2/\omega_f^2\right).
\end{equation}
Like in the QRM, we see that the $\pm$ branches within each doublet become degenerate as $\lambda_k\rightarrow \infty$ due to suppression by the exponential term $e^{-2\lambda_k^2/\omega_f^2}$ \cite{AA_1}. Different $k$ ladders however remain separated due to their distinct $-\lambda_k^2/\omega_f^2$ shifts. In addition, the intraband detunings in our model are seen to add energy-shift terms $\Delta^{\pm}_k$ that characterise the `average detuning' of the specific superpositions of superradiant states. 

\begin{figure}[h]
    \centering
    \includegraphics[width= 0.9\linewidth]{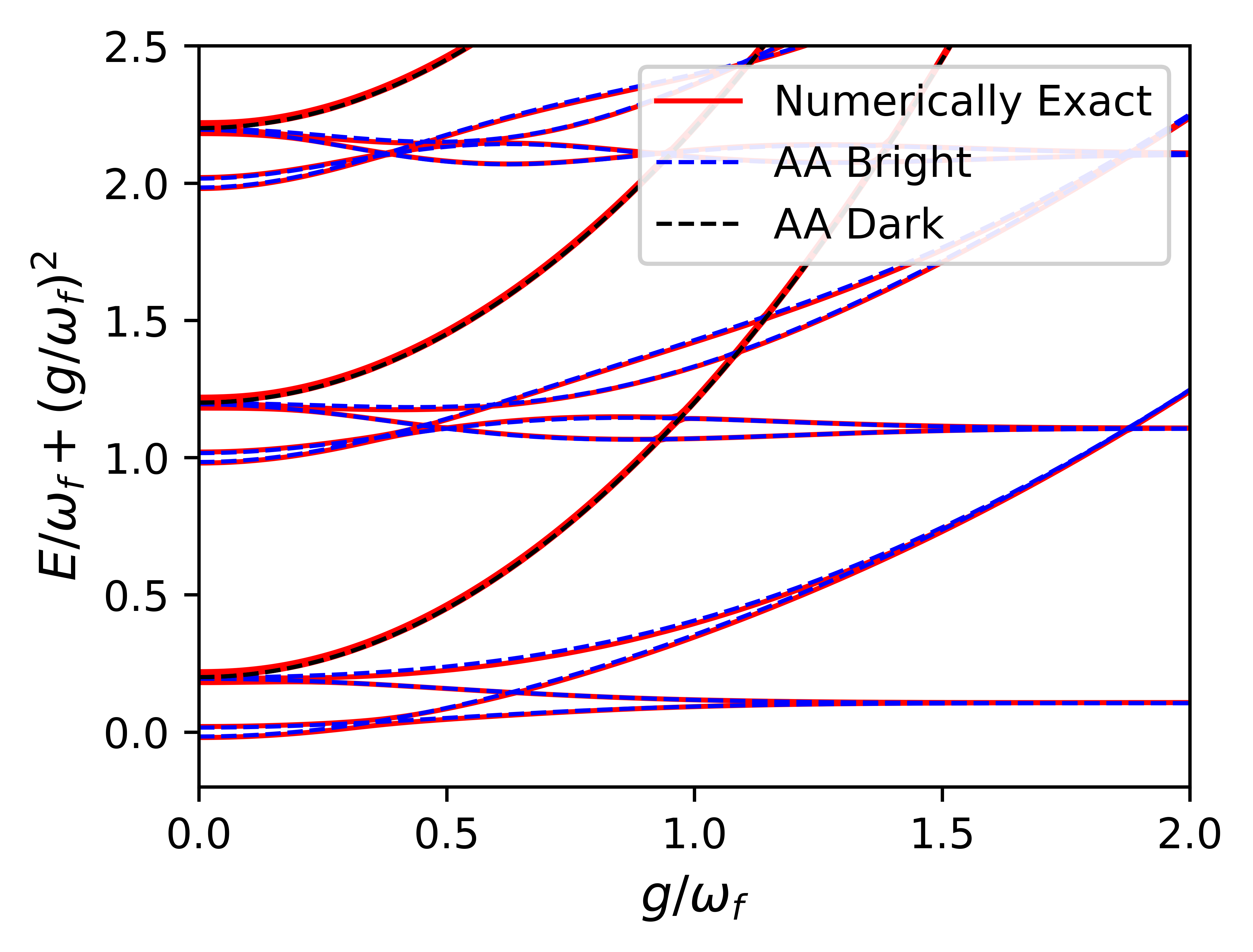}
    \caption{\textbf{Spectrum comparison between numerical diagonalisation and AA energies} for $D_g=2,D_e=4$ MQRM with $\omega_a=0.2\,\omega_f,\varepsilon=0.1\,\omega_a$, as a function of the coupling scale factor $g$. The maximum oscillator basis level for the numerical diagonalisation is chosen to be sufficiently high to avoid any truncation error ($N_\text{max} = 25$). Intraband detunings $\delta_k^{g/e}$ are specifically chosen as to be equally spaced within their respective domains, with $\boldsymbol \delta^g=[-1,1]$, and $\boldsymbol \delta^e=[-1,-1/3,1/3,1]$. The light--matter coupling matrix is drawn from a random distribution (see details in the main text). Our multilevel AA is in excellent agreement with the energy levels of the MQRM, while improving as $g/\omega_f$ increases.}
    \label{fig:AAverificationSPEC}
\end{figure}

In Fig.~\ref{fig:AAverificationSPEC}, we plot the numerical diagonalisation of a full $D_g=2,D_e=4$ MQRM Hamiltonian, with $M=2$ bright doublets and $N-M=2$ dark states, comparing it to its AA energies in order to verify the accuracy of our approximation. The light--matter coupling matrix is drawn from the complex normal distribution $\Lambda_{ij}\sim g~\mathcal{CN}(0,1)/{\rm[Max_{SV}}\mathcal{CN}(0,1)]$, where $g$ is simply a scale factor varied to change the overall coupling strength and ${\rm[Max_{SV}}\mathcal{CN}(0,1)]$ is the maximum singular value of $\boldsymbol{\Lambda}$ \cite{MQRM,RMT_Wishart_eigenvalues}. This choice of normalisation makes $\lambda_1=g$ by construction. We see that the AA recovers the energy levels of the full model, with expected inaccuracies near avoided crossings induced by intraband detunings not being evident at this scale. For small $g/\omega_f$ some minor inconsistency is noted, but this reduces overall as the light--matter coupling term dominates more strongly for increasing $g/\omega_f$. This is similar to the relationship between the original QRM and its corresponding AA energies \cite{AA_1,Braak2017,Xie2017}. 

\section{Thermal sensitivity}\label{sec:metro}

We consider the full light--matter MQRM as the thermometric probe. The probe is assumed to be brought into weak thermal contact with a sample or bath at temperature $T$ and allowed to equilibrate. Under the standard equilibrium-thermometry assumption that the probe--sample coupling thermalises the probe without appreciably renormalising its Hamiltonian, the equilibrated probe is well approximated by the Gibbs state of the MQRM Hamiltonian at the sample temperature, 
\begin{eqnarray}
\hat{\rho}_T = \frac{e^{-\beta\hat{H}}}{Z}, \quad\text{with }Z = \text{Tr } e^{-\beta\hat{H}},\quad \beta = 1/T.
\label{partitionfuncoriginal}
\end{eqnarray}
We set out to estimate the sample temperature $T$ from $\xi$ independent measurements on the probe. The statistical uncertainty of such estimates is tightly lower bounded by the Cramér--Rao bound and depends on the quantum Fisher information (QFI) $\mathcal{F}_T$ as \cite{PhysRevLett.72.3439}  
\begin{eqnarray}
    \mathrm{Var}[T]\ \ge\ \frac{1}{\xi\,\mathcal F_T(\hat\rho_T)}.
\end{eqnarray}

Since our probe is in the canonical Gibbs state, the optimal observable to estimate temperature is the system Hamiltonian $ \hat{H} $. The corresponding likelihood $ p(E\vert T) $ thus belongs to the exponential family and the Cramér--Rao bound becomes saturable even in a single shot ($\xi = 1$) \cite{kolodynski2014precision}. Thus, the QFI may be viewed as the ultimate indicator for the sensitivity of any thermometric scheme one may devise. For a probe in a Gibbs state, like ours, the QFI of the light--matter probe is proportional to its energy variance \cite{PhysRevE.83.011109,Therm_3_optimal}, i.e.,
\begin{eqnarray}
    \mathcal{F}_T(\hat{\rho}_T) = \frac{\langle \hat{H}^2\rangle- \langle \hat{H} \rangle^2}{T^4} = \frac{1}{T^4}\partial^2_\beta \ln Z(\beta).
    \label{QFIformula}
\end{eqnarray}
Equivalently, $\mathcal{F}_T = C(T)/T^2$, where $C(T)$ stands for heat capacity.
Direct evaluation of the QFI through diagonalisation of the full Hamiltonian is expensive in large Hilbert spaces. We therefore take advantage of our adiabatic approximation to obtain a closed-form expression for the QFI. This approximation is expected to be reliable in the adiabatic regime $\omega_a/\omega_f\ll 1$, while outside outside of it, neglected inter-manifold couplings may lead to sizable deviations in the spectrum and, therefore, shift the positions and heights of the QFI peaks. The AA provides us with a tractable partition function that can be used to analyse how the bright and dark manifolds contribute separately to the thermal response.

\begin{figure*}[th]
    \centering
    \includegraphics[width=0.75\linewidth]{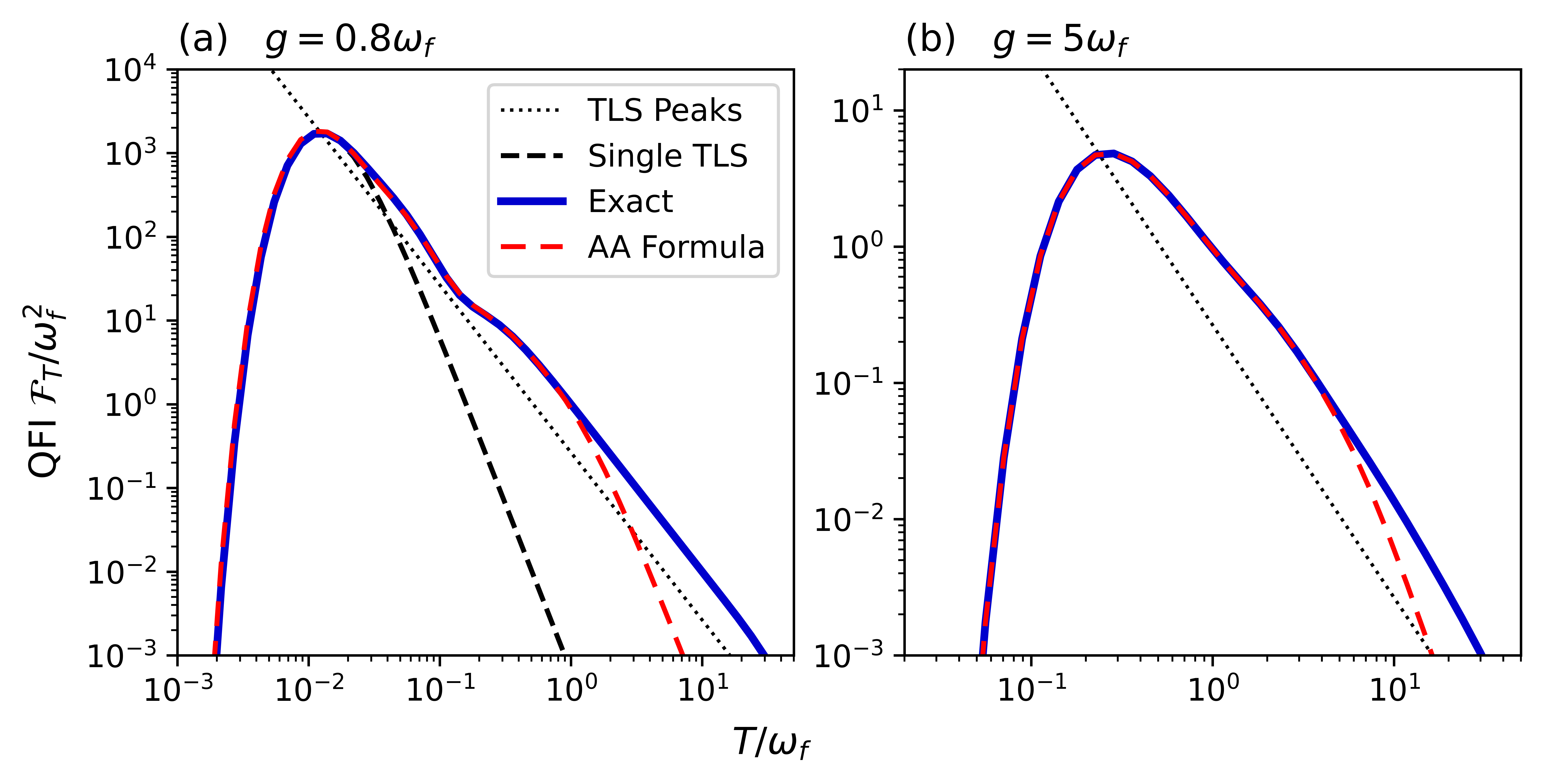}
    \caption{\textbf{Plots of AA QFI estimates and numerically exact calculations} from brute-force diagonalisation of a $D_g=2$, $D_e=4$ MQRM Hamiltonian. The truncation of the full MQRM Hamiltonian levels is chosen to be high enough to ensure convergence within the whole parameter regime. The lines tracing the locations of the peak QFI of TLSs optimally tuned for each temperature (cf. Eq.~\eqref{schottkyformula}) as well as the QFI of a fixed TLS resonant with the lowest bright doublet are plotted for comparison. In this case, they show that, over the temperature range of interest, the model has similar thermometric capability to the maximum of any simple two-level system, although over a much wider temperature range. System parameters are the same as in Fig.~\ref{fig:AAverificationSPEC}, including the light--matter couplings $\bL$. The `AA Formula' QFI is calculated from our Eq.~\eqref{QFI_formula}, and we retain terms that correspond to energies up to the 5-th dark manifold. \textbf{(a) Weak coupling regime, $g=0.8\omega_f$.} Despite the fairly large overall detunings $\varepsilon=0.1\,\omega_a$, the AA QFI recovers the exact result for the lowest temperatures, however, it underestimates the true value at higher temperatures due to truncation error. This can be mitigated by increasing the energy cutoff. \textbf{(b) Strong coupling regime, $g=5\omega_f$.} In the strong regime, the behaviour is very similar, with inaccuracy due to truncation at higher temperature. We otherwise succeed in recovering the exact QFI in the low temperature range.}
    \label{fig:AAverificationQFI}
\end{figure*}

Assuming that we are within the range of validity of the AA, such that $\omega_a/\omega_f \ll1$, we build the partition function
by adding the contributions from bright and dark manifolds (Eqs.~\eqref{AAbrightenergies} and ~\eqref{AAdarkenergies}) as
\begin{align}    \label{partitionfunc}
    Z(\beta) &= \sum_{n=0}^\infty\left[Z^B(n) + Z^D(n)\right] \\ 
    &= 2\sum_{n=0}^\infty\sum_{k=1}^M e^{\beta\,\gamma_B}\cosh{(\beta\,\Gamma_B)} + (N-M)\sum_{n=0}^\infty e^{\beta\,\gamma_D}, \nonumber
\end{align}
with
\begin{align}
    \gamma_B(n,k) &= \frac{\lambda_k^2}{\omega_f} - \frac{\Delta_k^+}{2}-\omega_f n, \nonumber \\
    \gamma_D(n,k) &= -\left(n\omega_f + \frac{\omega_a p }{2}\right),\\
    \Gamma_B(n,k) &= \frac{1}{2}(\omega_a+\Delta_k^-)e^{-2\,\lambda_k^2/\omega_f^2}\mathcal{L}_n\left(\frac{4\lambda_k^2}{\omega_f^2}\right). \nonumber
\end{align}
Since the QFI is invariant under global energy shifts, we drop the $+\omega_a/2$ offsets in Eqs.~\eqref{AAbrightenergies} and \eqref{AAdarkenergies}. The simplest way to proceed is by performing the derivative $\partial^2_\beta \ln Z$ from Eq.~\eqref{QFIformula} using a log-sum-exp approach \cite{logsumexp}. That is, we recast the partition function as some $Z(\beta) = \sum_i e^{x_i(\beta)}$, with $i$ running over every significant block (bright doublets $k$ as well as the dark manifold), with weights for each block given by $w_i(\beta) = e^{x_i(\beta)}/Z(\beta)$. The second derivative may then be cast as
\begin{align}
    \partial_\beta^2 \ln Z (\beta) &= \sum_i w_i(\beta)x_i''(\beta) \\&+ \left[\sum_i w_i(\beta)\left(x_i'(\beta)\right)^2-\left(\sum_iw_i(\beta)x_i'(\beta)\right)^2\right],\nonumber
\end{align}
or, equivalently, as $\partial_\beta^2 \ln Z (\beta) = \mathbb{E}_w[x_i''] + \text{Var}_w[x_i']$, where $\mathbb{E}_w$ denotes averaging over the weights $w$. Importantly, the structure of the partition function \eqref{partitionfunc} above means the expected value of $x_i''$ corresponds to the intrablock QFI contributions while the relevant variances bring in interblock contributions. Since every dark block is completely degenerate, the only contributions to the expected value part arise from the QFI contributions from each bright doublet. Taking the necessary derivatives, the QFI can be written explicitly as
\begin{eqnarray}
    \mathcal{F}_T = \frac{1}{T^4}\left(\frac{S_1(\beta)}{Z(\beta)} + \frac{S_2(\beta)}{Z(\beta)} - \frac{S_3(\beta)^2}{Z(\beta)^2}\right),
    \label{QFI_formula}
\end{eqnarray}
with
\begin{widetext}
\begin{subequations}
\label{eq:Si_functions}
\begin{align}
    S_1(\beta) &= 2\sum_{n=0}^\infty \sum_{k=1}^Me^{\beta\gamma_ B}\Gamma_B^2\text{ sech}\left(\beta\Gamma_B\right),\\
    S_2(\beta) &= \sum_{n=0}^\infty\left(S_2^B(n) + S_2^D(n)\right) =2\sum_{n=0}^\infty\sum_{k=1}^Me^{\beta\gamma_B}\cosh{(\beta \Gamma_B)}\left[\gamma_B + \Gamma_B \tanh{(\beta\Gamma_B)}\right]^2  +(N-M)\sum_{n=0}^\infty e^{\beta\gamma_D}\gamma_D^2,\\
    S_3(\beta)&=\sum_{n=0}^\infty\left(S_3^B(n) + S_3^D(n)\right)  = 2\sum_{n=0}^\infty\sum_{k=1}^Me^{\beta\gamma_B}\cosh{(\beta \Gamma_B)}\left[\gamma_B + \Gamma_B \tanh{(\beta\Gamma_B)}\right] +(N-M)\sum_{n=0}^\infty e^{\beta\gamma_D}\gamma_D,
\end{align}
\end{subequations}
\end{widetext}
where one can identify \cite{logsumexp}
\begin{eqnarray}
    \mathbb{E}_w[x_i'']=\frac{S_1(\beta)}{Z(\beta)}, \quad \text{Var}_w[x_i']= \frac{S_2(\beta)}{Z(\beta)} - \frac{S_3(\beta)^2}{Z(\beta)^2}.
\end{eqnarray}
In Appendix~\ref{app:qficomputing} below, we provide further details as to how to evaluate the terms in Eqs.~\eqref{eq:Si_functions} numerically by suitably truncating in $n$.

We now take the limit of the conventional QRM and consider only the lowest doublet, with $D_g=D_e=1$, $\lambda_1=g$. The truncation to $ n=0 $ reduces Eq.~\eqref{QFIformula} to the familiar expression for the QFI of a two-level system (TLS) with a renormalised level spacing of
\begin{eqnarray}
    \delta = \omega_ae^{-2g^2/\omega_f^2};
\end{eqnarray}
that is, the splitting $E_+^0-E_-^0 = \delta$, so that 
\begin{eqnarray}
    \mathcal{F}^{TLS}_{T} 
    = \frac{\delta^2}{4T^4}\text{sech}^2\left(\frac{\delta}{2T}\right).
    \label{twolevelsysQFIequation}
\end{eqnarray}

The behaviour of the TLS QFI is reminiscent of a Schottky anomaly in the heat capacity \cite{Gopal1966} in the sense that it peaks at some optimal temperature $ T^* $ related to its gap $\delta$ and decays rapidly both at higher and lower temperatures (exponentially fast in the low-temperature limit of $\beta\rightarrow\infty$ and as $\beta^2$ in the $\beta\rightarrow 0$ limit) \cite{Therm_Review}. 
We can derive the condition relating temperature $T^*$ and energy gap $\delta^*$ quite simply by setting $\partial_T\mathcal{F}_T^{TLS}=0$, which gives
\begin{eqnarray}
\delta/2T^* \approx 2.06534
\label{2ls_peaktotemp}
\end{eqnarray}
Hence, the function
\begin{eqnarray}
    \mathcal{F}^{TLS}_{T^*} = \frac{C}{T^{*2}}, \quad C\approx 0.265622
    \label{schottkyformula}
\end{eqnarray}
traces the point of maximum QFI of a simple TLS, 
providing a benchmark for thermal sensitivity. 

Beyond this TLS baseline, we use the same correspondence between dominant energy gaps and QFI maxima to interpret the thermometric enhancement achieved by our MQRM. In particular, saturating either its dark or bright sector amplifies interblock energy fluctuations and can produce responses that are both larger in magnitude and broader in temperature than a single TLS. This can already be seen on the high-temperature end of Fig.~\ref{fig:AAverificationQFI}, where we illustrate the accuracy of our QFI formula Eq.~\eqref{QFI_formula}. As we discuss in Sec.~\ref{sec:discussion} below, the `bump' in which sensitivity increases above the two-level reference is driven by the imbalance between $D_e$ and $D_g$, and remains remarkably robust to changes in the light--matter coupling terms. It is precisely this feature that we wish to characterise. 

\section{Examples and discussion}\label{sec:discussion}

\begin{figure}[h]
    \centering
    \includegraphics[width=0.9\linewidth]{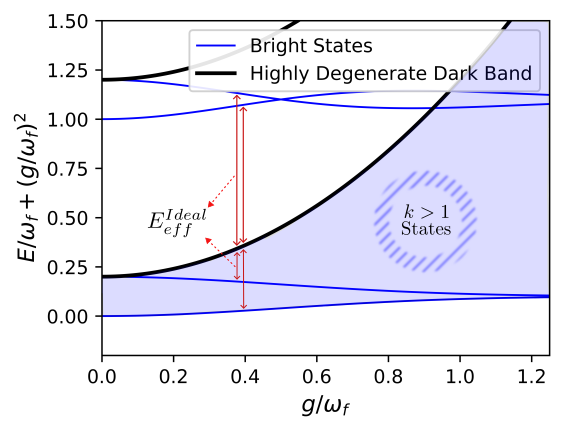}
    \caption{\textbf{Schematic energy spectrum of the MQRM with the lowest primary bright doublet} ($D_g=1$), and parameters chosen as $\omega_a=0.2\,\omega_f$, $\varepsilon=0$. 
    The solid red arrows indicate the thermometrically relevant transitions between bright and dark states at a given temperature, due to differing equilibrium-state populations. 
    For a given temperature that corresponds to a bright--dark peak of the MQRM QFI, these transitions define an effective energy gap $E_{\mathrm{eff}}^{\mathrm{ideal}}$ for a comparative benchmark ideal thermometer. We additionally show that any $n=0$ and $k>1$ bright doublets exist within the bounds of the ground state and the first dark band (shaded blue area).}
    \label{fig:QFIenergyscheme}
\end{figure}

We now probe the thermometric performance of the MQRM in cases in which the atomic manifold is large and highly structured. 
As we shall see, the light--matter couplings and the allocation of levels between bright and dark sectors are crucial in determining the maximum sensitivity and the temperature range of effective operation. Specifically, we consider two complementary situations (cf. Fig.~\ref{fig:QFIenergyscheme}). In the first, which we term `dark-manifold saturation', the atom has relatively few nondegenerate ground states coupled via a cavity mode to a large, nearly degenerate, excited manifold. In that case, we show how the interplay between the bright states and the dark manifold can generate QFI peaks that outperform our reference temperature-optimised two-level probes (cf. the high-temperature QFI peak shown in Fig.~\ref{fig:AAverificationQFI}). This type of configuration closely mirrors the level structure of semiconductor quantum dots, where the band edge exciton exhibits a high degree of degeneracy \cite{Review_USC_Polariton, PbSorSeQDpaperoriginal,64degenQD,64DegenQD2,TommasoQRMExperimental}. In this sense, the MQRM under dark manifold saturation can be viewed as a minimal cavity-QED abstraction of such platforms, while neglecting the detailed fine structure and dissipation. 

In the second situation that we consider, the configuration of `bright manifold saturation', the number of bright doublets is maximised by choosing $D_g\approx D_e$ so that most atomic superpositions couple appreciably to the cavity, while the truly dark subspace is comparatively small. This situation is representative of multilevel systems in the ultrastrong coupling regime, where several transitions hybridise with a confined mode, such as inter–Landau-level cyclotron transitions in cavity-embedded two-dimensional electron gases \cite{Review_USC_Polariton,USC_Polariton,GaAsQW,InAsQW,HeliumQW}. Here, we retain only the generic feature of having multiple bright transitions with different effective couplings and leave out complications, such as detailed selection rules and multimode photonic structure.

For dark-manifold saturation, the natural benchmark is the ideal thermometer \cite{reeb2015tight,Therm_3_optimal}, with parameters chosen to match the effective energy gap and degeneracy of our MQRM for direct comparison. In turn, for bright manifold saturation, the focus is on typical behaviour under random light--matter couplings, and on how the subsequent most likely distribution of singular values shapes both the height and width of the QFI profile. Finally, we also use these examples to assess thermometric sensitivity when global access to the full light--matter probe is not practically feasible, by comparing the global QFI with the QFI obtainable from measurements on the reduced atomic and cavity states.

\subsection{Dark manifold saturation}

We consider an atom with $D_g$ ground states and $D_e$ excited states so that, in the SVD basis, the model contains $D_g$ bright doublets and $D=D_e-D_g$ dark states, with $p=1$ indicating that the excess states lie within the excited band (cf. Eq.~\eqref{eq:dark-Hamiltonian}). The light--matter coupling is normalised as in Sec.~\ref{AA_section}, and we denote the largest singular value as $\lambda_1 = g$. The remaining singular values of the coupling matrix $\boldsymbol{\Lambda}$ are fixed using the standard Laguerre–Wishart construction, detailed in Appendix~\ref{laguerreappendix}. There, we provide a self contained derivation for the most likely SVD spectrum for a Ginibre-ensemble random coupling matrix, with elements iid sampled from a complex normal distribution $\Lambda_{ij}\sim \mathcal{CN}(0,1)$ \cite{LagWish_1,LagWish_2,Lagwish_3}. In the dark-manifold regime of interest, $D_g$ is kept small while $D$ is increased, so that the number of dark excited states becomes large at fixed bright structure and coupling scale.

\subsubsection{Effective energy gap of a comparative ideal thermometer}

An expression for the bulk of the QFI at the peak mediated by the bright--dark manifold interaction, valid when $D \gg D_g$, follows directly from the general QFI formula Eq.~\eqref{QFI_formula}. Specifically, retaining only the bright--dark interblock terms the relevant QFI would reduce to
\begin{equation}
    \mathcal{F}_T\approx \mathcal{F}_T^{BD} = \frac{\beta^4}{Z^2}
    \bigl(S_2^B Z^D + S_2^D Z^B - 2 S_3^B S_3^D\bigr),
\label{darkqfi}
\end{equation}
where $Z^{B/D}$ and $S_{2,3}^{B/D}$ denote, respectively, the bright and dark contributions to the partition function and to the moments entering Eq.~\eqref{QFI_formula}. We find that, in the dark-manifold saturated regime, it is precisely this contribution to the total QFI which captures the high-temperature feature surpassing the two-level benchmark sensitivity shown in Fig.~\ref{fig:AAverificationQFI} (see also Fig.~\ref{fig:peakratio_heatmap} below). 

This can be understood directly from the adiabatic spectrum depicted in Fig.~\ref{fig:QFIenergyscheme}. We know that the lowest few excitations above the ground state are bright doublets with varying levels of light--matter coupling. The dark states, by contrast, appear as the highest-lying levels within the $n=0$ manifold. Hence, transitions between the two will become activated at comparatively high temperatures.
In turn, the bright--bright contributions to the QFI primarily shape the low-temperature intra-doublet bright peak and the intermediate inter-doublet structure (cf. Fig.~\ref{fig:peakratio_heatmap}). In general, weaker light--matter couplings can give rise to several overlapping contributions peaking within the same temperature range, 
making it difficult to isolate the bright--dark contribution. However, in the limit $D\gg D_g$, we have found the high-temperature `bump' of the QFI profile is accurately reproduced by $\mathcal{F}_T^{BD}$ alone, due to the substantial difference in magnitude between the various contributions. 

\begin{figure*}[th]
\centering
\includegraphics[width=0.45\linewidth]{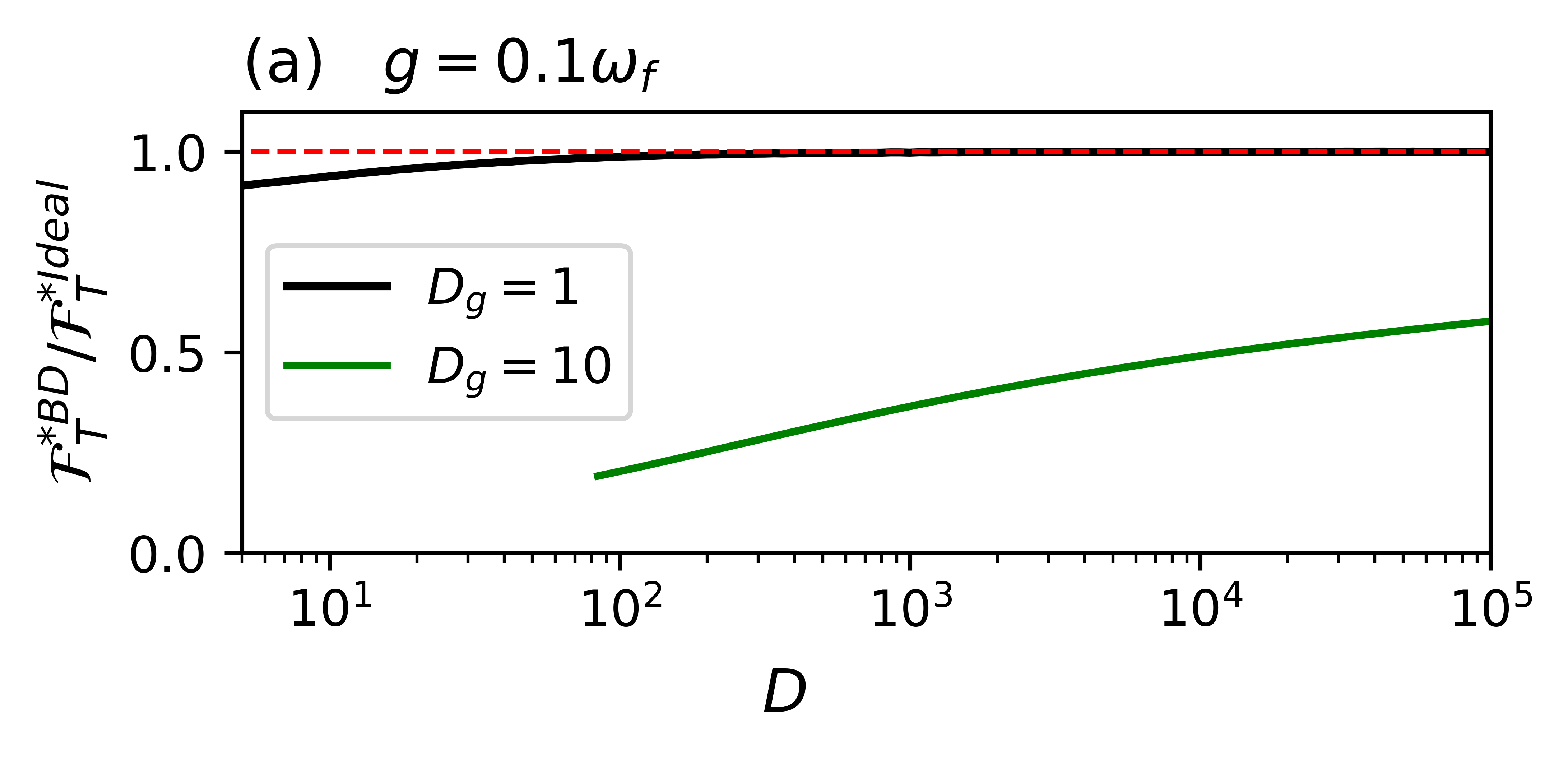} \quad 
\includegraphics[width=0.45\linewidth]{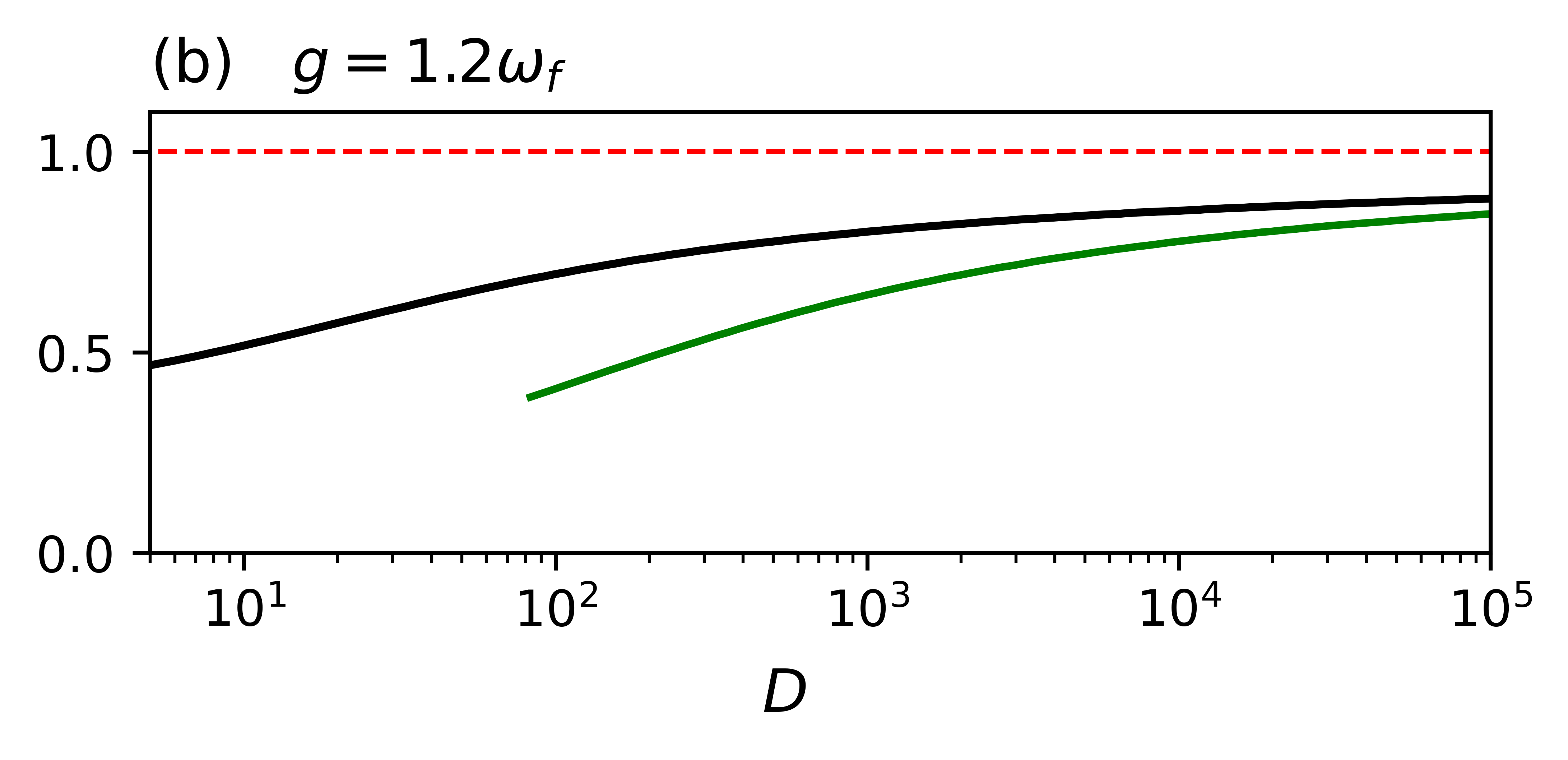}
\caption{\textbf{Dark manifold saturation regime and the approach towards the ideal thermometer. (a)~Weak--coupling regime $g=0.1\omega_f$. (b)~Intermediate coupling $g=1.2\omega_f$}. Atomic frequency is chosen as $\omega_a=0.2\omega_f$, with $\varepsilon=0$ intraband detuning. The figures show the ratio between the maximal QFI of the dark-bright manifold MQRM and that of the corresponding ideal probe, $\mathcal{F}_T^{*\,\text{BD}}/\mathcal{F}_T^{*\,\text{ideal}}$, as a function of the dark state degeneracy $D$ for different numbers of ground states $D\gg D_g$ (black: $D_g=1$, green: $D_g=10$). The red dashed line marks the ideal limit $\mathcal{F}_T^{*\,\text{BD}}/\mathcal{F}_T^{*\,\text{ideal}}=1$. We see that, for all $D_g$, increasing the size of the dark manifold systematically raises the peak-QFI ratio $\mathcal{F}_T^{*\,\text{BD}}/\mathcal{F}_T^{*\,\text{ideal}}$ and drives the MQRM towards ideal thermometric behaviour in both the weak and intermediate coupling regimes. In the weak coupling regime the bright doublets compete for thermal weight and, as we would expect, strongly reduce the fraction of population that participates in the desired dominant bright--dark process. This hinders the overall sensitivity. Note also how the atomically ideal but light--matter coupled $D_g=1$ configuration becomes comparatively less efficient. This is due to the fact that stronger couplings populate higher primary bright ladder states, weakening the effective bright--dark transition. In turn, for $D_g=10$, the trend is reversed, with the peak QFI ratio improving slightly with increased coupling. This indicates that additional bright ladders can compensate for the stronger coupling in the intermediate coupling regime.
}
\label{fig:peakratioDark1D}
\end{figure*}

We assess the performance of our dark-manifold saturated MQRM probe at its high-temperature peak by comparing it to an ideal thermometer, with a single nondegenerate ground state and a $D$-fold degenerate excited manifold, calibrated for maximum sensitivity at the peak temperature. The thermal QFI of such probe is discussed in Appendix~\ref{idealquantumtherm_appendix}. Since the full bright--dark MQRM QFI expression $\mathcal{F}_T^{BD}$ is too cumbersome to invert analytically, we determine the optimal gap $E_\text{eff}^\text{ideal}$ for the ideal thermometer numerically, by imposing that the ideal-probe QFI \cite{Therm_3_optimal} reaches its maximum at the same temperature as $\mathcal{F}_T^{BD}$ with the same dark-state degeneracy. This fixes a like-for-like comparison. 
The procedure is as follows. First, we compute $\mathcal{F}_T^{BD}$ numerically and identify the peak, $\mathcal{F}_T^{*\,BD}$, occurring at $T^*$. Then, for the ideal thermometer with upper degeneracy $D$, the condition for the peak can be written in terms of the renormalised temperature $x = \beta E$ as
\begin{eqnarray}
    x = \ln\left(D\frac{x + 4}{x - 4}\right).
\end{eqnarray}
For fixed $D$ this equation has a single relevant solution with $x>4$. Using $\beta^*=1/T^*$, we can then calculate $E_\text{eff}^\text{ideal} = x \, T^*$, revealing the effective gap within a dimensionally comparative ideal system. 

\subsubsection{Peak scaling and ensemble stability}

\begin{figure*}[th]
\centering
\includegraphics[width=0.47\linewidth]{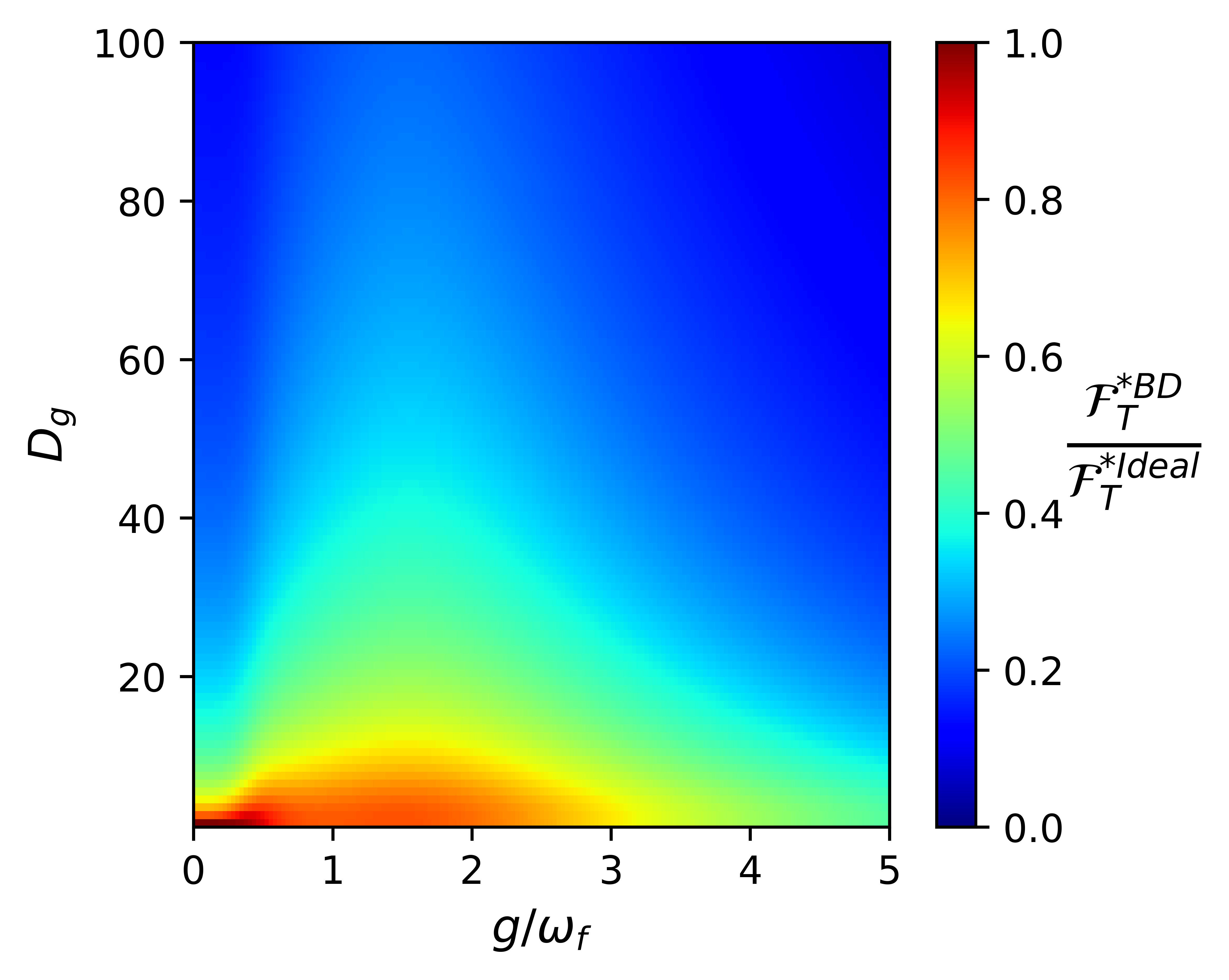}
\includegraphics[width=0.47\linewidth]{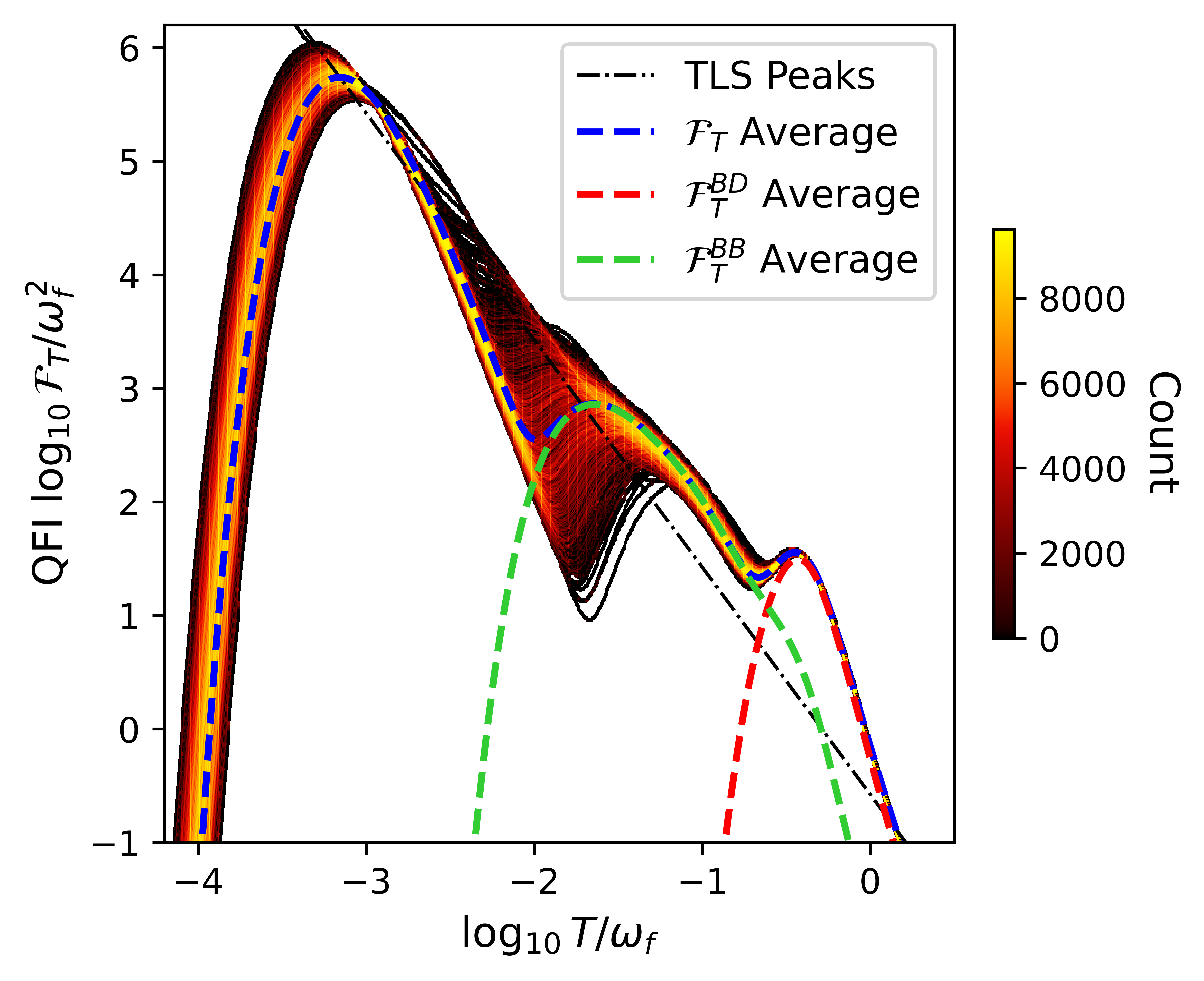}
\caption{\textbf{Dark manifold saturated thermometry in larger systems. Left: Ratio between the maximum bright--dark contribution to the QFI of a MQRM and that of the corresponding ideal probe}, $\mathcal{F}_T^{*\,BD}/\mathcal{F}_T^{*\,\text{ideal}}$, for a fixed dark state degeneracy $D=1000$ as a function of lower manifold degeneracy $D_g$ and coupling $g/\omega_f$ ($\omega_a=0.2\,\omega_f$ and $\varepsilon=0$). For a heavily dark--saturated atom the ratio is non-monotonic with respect to the coupling. For small $D_g$ the peak QFI improves when the coupling is increased from the weak to the intermediate coupling regime, while at stronger coupling, the performance degrades as higher bright ladders become populated. \textbf{Right: Density heatmap of dimensionless QFI curves for an ensemble of $1\times 10^4$ random MQRM realisations} with $D_g=10$ and $D=1000$ ($\omega_a=0.25\omega_f$, $\varepsilon=0.25\omega_a$ and intraband detunings uniformly distributed as $\delta^{e/g}_i\in[-1,1]$). The `count' density represents the number of lines within a single pixel of the plot. The coupling matrices are drawn from the complex Ginibre ensemble and rescaled such that the average largest singular value is $g\approx1.5\omega_f$, so that the systems are in the intermediate coupling regime. The black dash-dotted line marks all traced two-level peaks, while the blue, red, and green dashed lines show respectively, the typical disorder-free realisations of the total QFI, its bright--dark contributions and its bright--bright contributions. Despite great disorder in detunings and couplings, the QFI remains stable around the average, with the near-ideal high-temperature dark peak associated with bright--dark transitions being particularly robust.}
\label{fig:peakratio_heatmap}
\end{figure*}

We now look at the peak thermometric performance of the MQRM, relative to the benchmark ideal probe. Fig.~\ref{fig:peakratioDark1D} shows how increasing the size of the dark manifold pushes the MQRM towards the ideal limit for an atom with and without multiple bright doublets due to the dominant bright--dark process. For all couplings and all $D_g$ we see that increasing the size of the dark manifold brings the QFI closer to the ultimate limit set by the optimal probe. 
 
 

For Fig.~\ref{fig:peakratioDark1D}, the singular values of the coupling matrix were chosen in a way that neatly separates the effect of dark-state degeneracy from changes in the bright sector. For each fixed $D_g$ we determined the `typical' set of singular values $\{\lambda_k\}$ using the machinery in Appendix~\ref{laguerreappendix} (see also Refs.~\cite{LagWish_1,LagWish_2,Lagwish_3}) for a square $D_g\times D_g$ complex Ginibre matrix. This is the set of singular values one would get on average if all states were bright. These $\lambda_k$ (with $\lambda_1=g$) are then held fixed as $D$ is increased, and the additional entries in the coupling matrix are taken to be dark. In this way, the bright couplings do not depend on $D$ and any changes in Fig.~\ref{fig:peakratioDark1D} can be completely attributed to the growth of the dark manifold, rather than to modifications on the bright sector secondary doublet couplings. Nonetheless, we have checked that performing instead a `true average' so that the singular values depend on $D$ does not qualitatively change any of our results.

The left-hand panel of Fig.~\ref{fig:peakratio_heatmap} neatly illustrates the role of the light--matter coupling in the optimality of a MQRM in the dark-saturated regime, when given the presence of a small number of bright doublets. We see that for such an atom, the optimal ratio increases in the intermediate coupling regime.
This can be traced back to a restructuring of the spectrum as the overall coupling increases, resulting in the highly degenerate dark band getting pushed away from the upper bright ladder states (see Fig.~\ref{fig:QFIenergyscheme}). For intermediate couplings, this effect isolates a single dominant bright--dark gap,
which enhances the associated QFI peak ratio. We do not expect, however, such intermediate-coupling QFI boost to be universal, as it must rely on how the secondary bright doublet scales with $g/\omega_f$ and thus, on the singular value spectrum of the coupling matrix. For the most likely set of singular values used in Fig.~\ref{fig:peakratioDark1D} (left), we have a relatively even distribution, and so the bright--dark gap is optimally isolated at intermediate $g/\omega_f$. For more unbalanced singular value statistics, the reshuffling could be less favourable and the enhancement might be weakened.

We also study how different parts of the spectrum contribute to the overall QFI in Fig.~\ref{fig:peakratio_heatmap}. Namely, we realise a large ensemble of highly degenerate, disordered systems with an average primary coupling of $g=1.5\,\omega_f$. The `average' QFI curves plotted are not ensemble averages, but are obtained by evaluating the analytic QFI formulas for a single MQRM with vanishing detunings and a coupling matrix with singular values set to their most likely values. In this sense, the averages represent a typical, disorder-free realisation of the total ensemble, against which the disordered randomised ensemble can be compared. The bright--dark curve is calculated from the corresponding QFI contribution of Eq.~\eqref{darkqfi}, and the bright--bright manifold contribution can be extracted similarly, with
\begin{eqnarray}
    \mathcal{F}_T^{BB} = \frac{\beta^4}{Z^2}\left(S_2^B Z^B - [S_3^B]^2\right).
    \label{brightbright}
\end{eqnarray}

We could additionally extract the contribution corresponding to fluctuations between different dark manifolds, but as these are each degenerate and periodic in energy the QFI response would be that of a typical Harmonic Oscillator and therefore contribute only little at higher temperatures. 
At the lowest temperatures, the peak in the QFI is well captured by the two-level system peak guide line and can thus be attributed to the tiny splitting between the lowest bright-doublet levels. The intermediate temperature peak is accounted for by the bright--bright manifold contribution, and corresponds to transitions between bright blocks and, again, is captured well by the two-level guide. For random systems, the high variance in secondary-to-primary singular value ratios of the randomly generated coupling matrices is responsible for the largest variations. At the highest temperature, the QFI is dominated by the bright--dark contribution, which reflects transitions between the bright sector and the densely populated dark manifold, and gives rise to a near-ideal QFI contribution orders of magnitude larger than the two-level guide. The total QFI is a combination of all these features, with a useful temperature range of effective operation much larger than that of a perfect ideal thermometer, but with the bright--dark maximum contribution attaining a comparable peak sensitivity. Importantly, Fig.~\ref{fig:peakratio_heatmap} also shows that the performance of our MQRM probe is very 
robust against microscopic disorder in individual coupling matrix elements and the intraband detunings.

\subsection{Bright manifold saturation}

\begin{figure*}[th]
\centering
\includegraphics[width=0.47\linewidth]{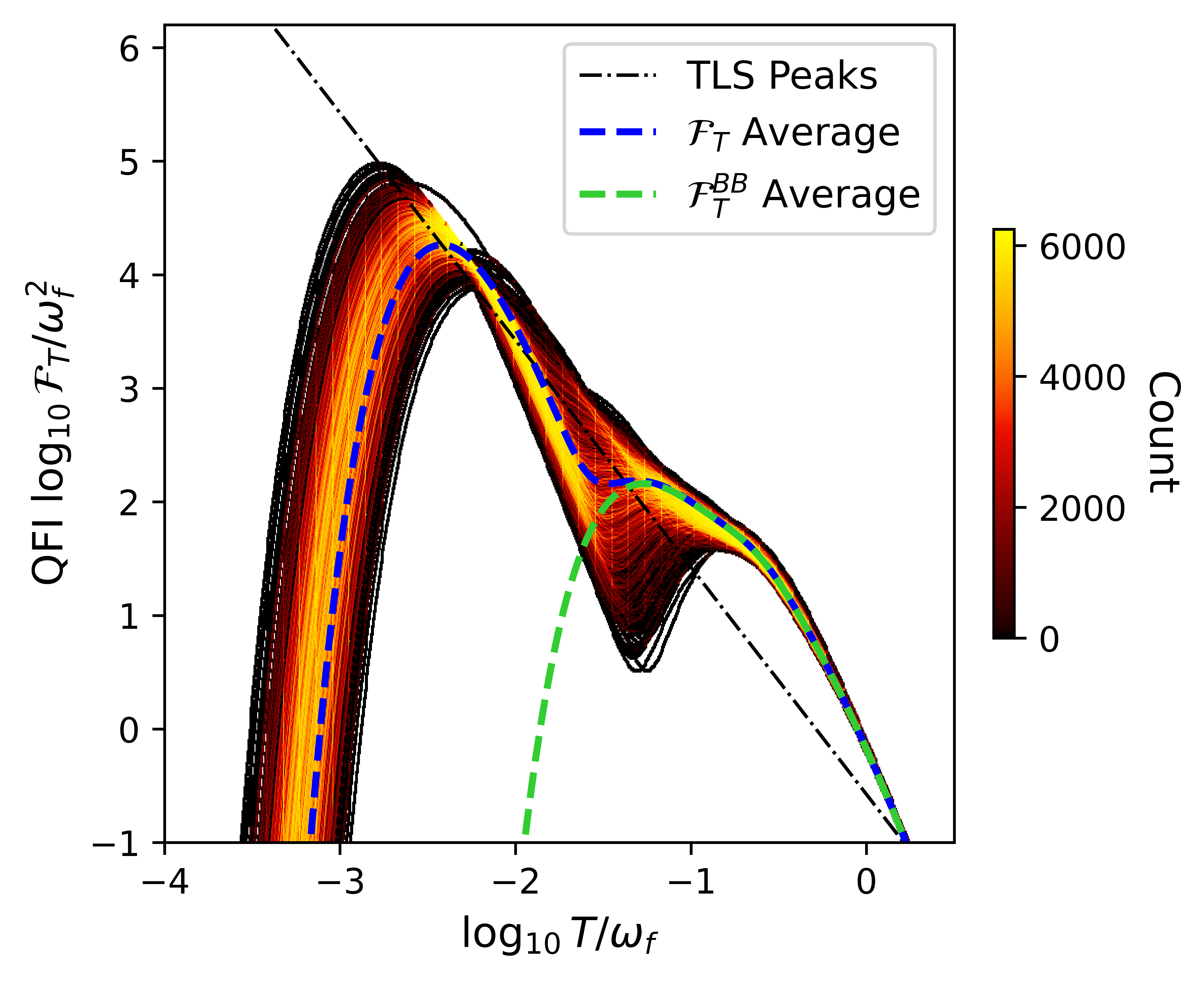}
\includegraphics[width=0.47\linewidth]{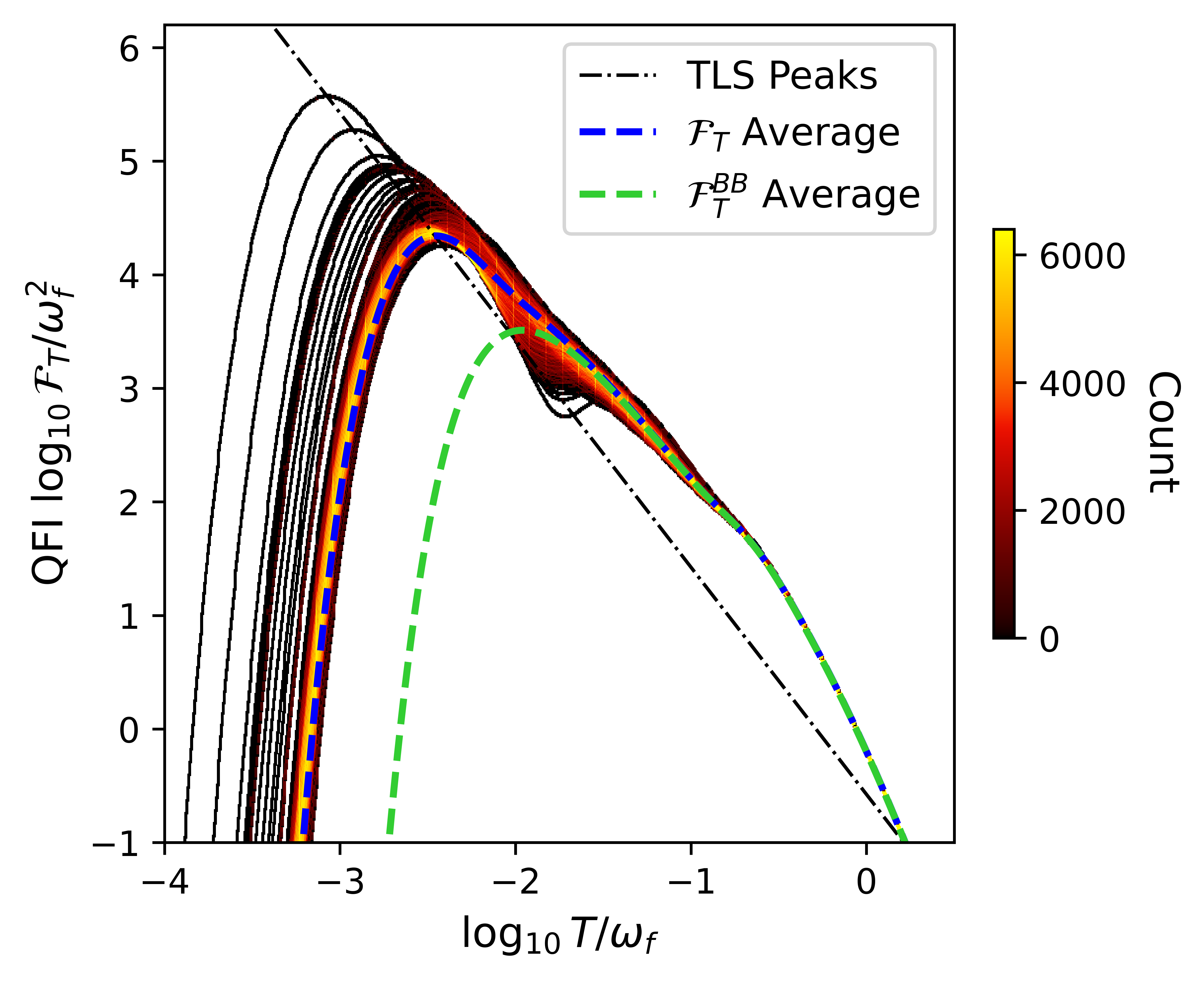}
\caption{\textbf{Bright manifold saturation and manifold size induced self averaging of the thermometric response at intermediate coupling.} Density heatmaps of QFI curves for ensembles of random MQRM realisations with fully bright atomic manifolds $D_g=D_e$, $\omega_a=0.25\omega_f,\varepsilon=0.25\,\omega_a$, and intraband detunings uniformly distributed as $\delta^{e/g}_i\in[-1,1]$. \textbf{(left) $M=25$}, with $8000$ realisations, and \textbf{(right) $M=250$}, with $6500$ realisations. The `count' density represents the number of lines within a single pixel of the plot. The coupling matrices are drawn from the complex Ginibre ensemble and rescaled such that the average largest singular value is $g\approx1.2\,\omega_f$. The black dash-dotted line traces the peak locations of an effective two-level system for comparison. The blue dashed curve shows the average QFI obtained by evaluating the analytic expressions with vanishing detunings and modal singular values from the Laguerre--Wishart construction, while the green dashed curve shows the corresponding bright--bright contribution $\mathcal{F}_T^{BB}$. Increasing the manifold size broadens the bright--bright features and simultaneously reduces sample-to-sample variability, with QFI curves concentrating more tightly around the typical response. This indicates a degree of self averaging and a progressively more stabilised broadband thermometric window for larger manifolds.}
\label{fig:brightsatMC}
\end{figure*}

\begin{figure*}[th]
\centering
\includegraphics[width=0.8\linewidth]{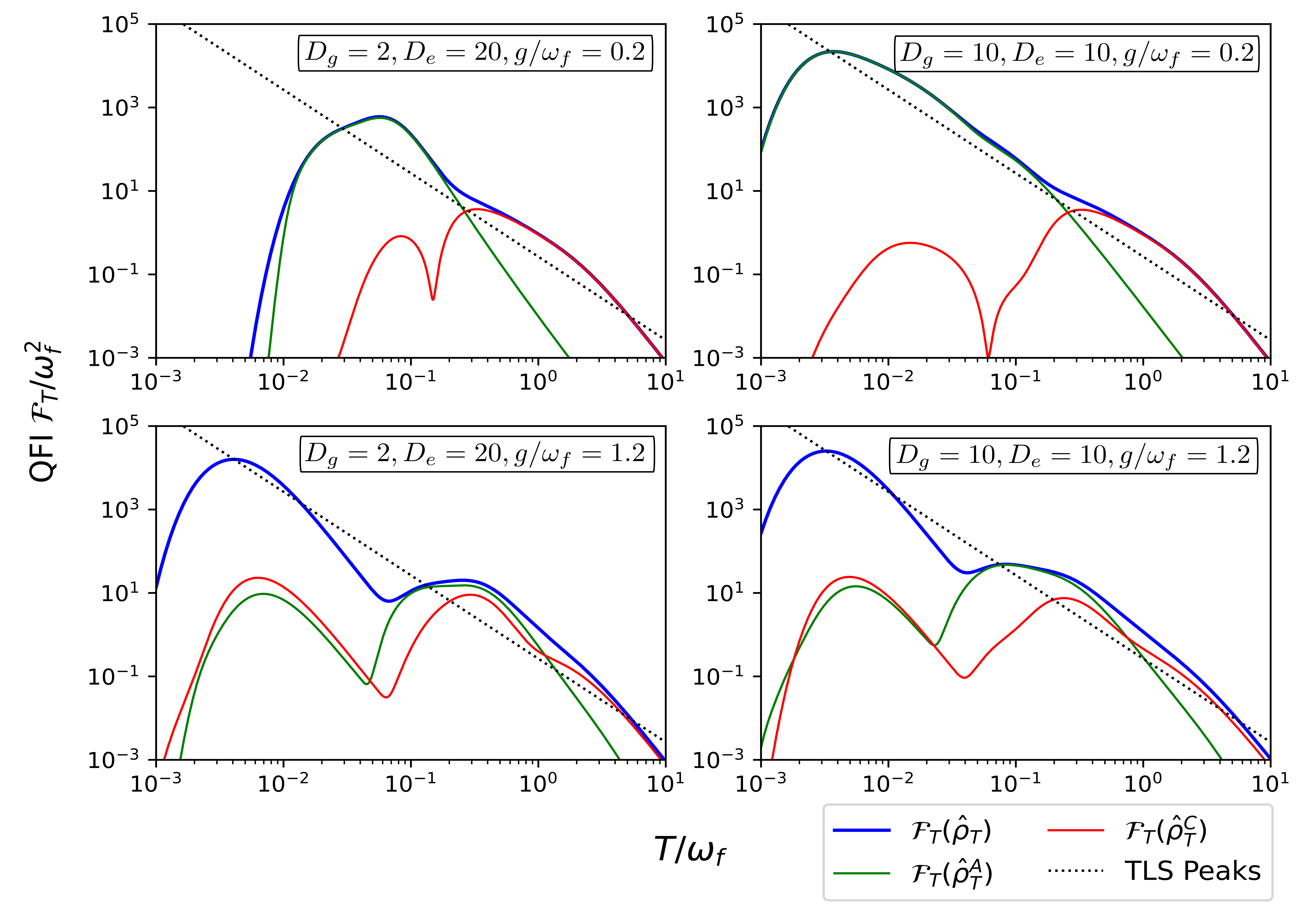}
\caption{\textbf{Comparison between global and reduced-state thermometric sensitivity in the MQRM.} QFI curves are shown for the full light--matter Gibbs state $\mathcal{F_T(\hat \rho_T)}$ in blue, the reduced atomic state $\mathcal F_T(\hat\rho_T^A)$ in green, and the reduced cavity state $\mathcal F_T(\hat\rho_T^C)$ in red, with the optimised TLS benchmark shown for reference. The left panels correspond to dark saturation, with $D_g=2,D_e=20$, while the right panels correspond to bright-manifold saturation, with $D_g=D_e=10$. The upper and lower panels compare weaker and stronger light--matter coupling, respectively. In the temperature regions where the global QFI is enhanced by the multilevel bright--dark structure, access to either reduced subsystem retains a substantial fraction of the full thermometric sensitivity, showing that the multilevel enhancement is not entirely dependent on ideal global measurements of the complete light--matter probe.}
\label{fig:refplots}
\end{figure*}

We now discuss the situation in which the number of bright doublets is maximised by taking $D_g=D_e$, together with a full rank coupling matrix. In the SVD basis this produces $M$ bright doublets and $N=0$ dark states, so the resulting thermometric response is entirely due to bright--bright processes $\mathcal{F}_T^{BB}$ (cf. Eq.~\eqref{brightbright}). Unlike dark-manifold saturation, where a single highly degenerate bright--dark channel dominates the response,
 bright saturation distributes thermometric sensitivity across many distinct bright transitions. Within the adiabatic limit, this appears as a set of bright ladders whose energies `fan out' with the singular values $\{\lambda_k\}$, generating not just many individual doublet splittings, but a dense array of inter-doublet energy differences belonging to different $k$-ladders. These numerous bright--bright gaps populate a wide range of energy scales so, regardless of temperature, there is always a sizeable set of thermally active transitions contributing to the QFI. The resulting QFI is, therefore, broad by construction and
 can yield extended temperature intervals that outperform the TLS guide Eq.~\eqref{schottkyformula}. 

The detailed shape and bandwidth of the $\mathcal{F}_T^{BB}$ response is controlled primarily by the singular value spectrum, whereby broad and relatively even $\{ \lambda_k\}$ distributions will produce a dense and well spread gap structure and therefore a robust plateau. A highly structured spectrum, on the other hand, could reduce the overall energy variance and drive the response back to near-TLS behaviour with a single well-defined peak. For example, if the singular values are all degenerate, with $\lambda_k=g$ for all $k$, then the bright ladders no longer fan out and, instead, acquire essentially the same displacement and spacing. In that case, the spectrum effectively reduces to repeated copies of the same QRM structure with multiplicities, and the thermometric response loses its broadband enhancement. 

Here, we focus on the generic case of random light--matter couplings, drawing $\bL$ from an iid complex Ginibre ensemble, for which the modal singular value spectrum is well described by the Laguerre--Wishart construction discussed in Appendix~\ref{laguerreappendix}. This choice provides a full-rank coupling profile with a broad set of $\{\lambda_k\}$, therefore capturing the `typical' bright saturation behaviour without relying on fine tuning.

In Figure~\ref{fig:brightsatMC} (left), we show the bright-saturated response heatmap for a moderate manifold size $M=25$. The ensemble produces a broad QFI profile dominated by bright--bright processes, with multiple local maxima and broad peaks, rather than a single isolated feature. The peak at $\log_{10} T \approx -2.5$ can be attributed to the primary bright--bright intra-doublet spacing. 
At higher temperatures the response is instead governed by transitions between different bright ladders. In this regime, the total QFI is well approximated by the bright--bright sector expression $\mathcal{F}_T \approx \mathcal{F}_T^{BB}$, and the enhancement above the TLS guide is distributed over a finite temperature interval rather than a concentrated single peak. 

Figure~\ref{fig:brightsatMC} (right) shows the heatmap after increasing the manifold size to $M=250$.
The relevant energy scales entering $\mathcal{F}_T^{BB}$ are the bright--bright transition energies, which are controlled by differences between the many ladder energies and, therefore, depend on the full set of pairwise combinations of $\{\lambda_k\}$. Increasing $M$ from $25$ to $250$ dramatically increases both the number and the density of these gap scales, so the thermal activation as a function of $T$ is spread over a wider interval, and the resulting QFI feature broadens towards lower temperatures and becomes smoother. At the same time, the sample-to-sample variance decreases for large $M$, which is due to two separate effects. Firstly, the QFI becomes increasingly self-averaging because it is the sum of many comparable bright--bright contributions. Hence, fluctuations in any individual singular value or small subset of channels has a diminishing effect on the total energy variance. Also, for large random matrices, the fluctuations of the ordered singular values themselves shrink with matrix size, so the spectrum $\{\lambda_k\}$ effectively concentrates around its typical (Laguerre--Wishart) profile. 

\subsection{Limited access to the probe}

Saturating the global QFI requires optimal measurements on the full joint light--matter state. In practice, however, experimental access may be restricted to a particular subsystem (i.e., either the atomic or the cavity marginal).
We, therefore, ask the question of how much of the global thermometric information may be extracted locally. For the full MQRM Gibbs state $\hat \rho_T$, we define the reduced atomic and cavity states by 
\begin{equation}
    \hat \rho^A_T = \Tr_C[\hat \rho_T],\qquad \hat \rho^C_T = \Tr_A[\hat \rho_T],
\end{equation}
 where $A$ denotes the multilevel atom and $C$ denotes the cavity. These reduced states are generally not Gibbs states of the bare atomic or cavity Hamiltonians. Consequently, their QFIs cannot be computed from a local energy variance and, instead, we evaluate the full mixed-state QFI of the reduced density matrix. For a reduced state $\hat \rho^X_T$, with $X\in \{A,C\}$, we use \cite{QFI_review}
 \begin{equation}
\mathcal F_T[\hat\rho^X] =
2\sum_{\mu,\nu} \frac{ \left| \bra{\mu} \partial_T \hat\rho^X \ket{\nu} \right|^2 }{ p_\mu+p_\nu },
\label{eq:reduced_state_qfi}
\end{equation}
where $\hat \rho^X = \sum_\mu p_\mu \ket{\mu}\bra{\mu}$, and terms with $p_\mu + p_\nu = 0$ are ommitted. 

Numerical results are shown in Fig.~\ref{fig:refplots} for relatively small manifold sizes, with representative examples of both dark-saturated and bright-saturated MQRM thermometry. As expected from monotonicity of the QFI under partial trace, the reduced QFIs remain bounded above by the global QFI. Nevertheless, the atomic reduced state preserves a large fraction of the global response over much of the temperature range shown. The main exception is the lowest-temperature peak visible in the $g=1.2\,\omega_f$ panels. This feature is (as before) associated with the lowest Rabi-like bright doublet, and is thus essentially a two-level contribution, rather than a genuinely multilevel enhancement. Its partial loss when considering local QFIs is consistent with the strong atom--cavity hybridisation of this primary bright doublet, where tracing out the cavity removes a significant part of the relevant temperature-dependent structure.

The preservation of the broader multilevel response, especially in the reduced atomic QFI, follows from the remaining bright--dark structure of the MQRM. In the adiabatic picture, the $k$-th bright sector is associated with displaced oscillator states $\hat D(\mp \lambda_k/\omega_f)\ket n$, so the degree of atom--cavity hybridisation is controlled by the individual doublet singular values $\lambda_k$. The primary bright doublet, with the largest coupling, is therefore the most strongly displaced, and tracing out the cavity can remove a substantial part of its temperature-dependent structure. By contrast, subdominant bright and near-dark sectors have smaller couplings $\lambda_k$, so their displaced oscillator states are less separated and their reduced states retain more of the thermal population sensitivity responsible for the global QFI. This retention should become stronger as the manifold sizes increase, since the singular-value spectrum then contains more weakly coupled subdominant sectors in addition to the primary bright mode.

The cavity reduced QFI is generally less effective at recovering low-temperature features. These features are associated with small gaps in weakly hybridised or primarily atomic sectors, so tracing out the atom removes much of the structure that distinguishes the relevant thermal populations. At higher temperatures, more oscillator manifolds become populated, and changes in the displaced cavity states become more visible in the reduced cavity density matrix. The cavity reduced QFI therefore tends to perform better at higher temperatures.

Overall, Fig.~\ref{fig:refplots} shows that the near-ideal and broad MQRM thermometric response can be attained even when measurements on the probe are limited by local access.
These sectors retain more of their temperature-dependent structure after partial tracing, so they provide local thermometric information even when stronger bright modes are more strongly hybridised with the cavity. Determining how large the manifolds must be to guarantee a given fraction of the global QFI for a given primary coupling scale is a separate scaling question, which we leave for future work.

\section{Conclusion}\label{sec:conclusion}

We have analysed equilibrium thermometry in a multilevel quantum Rabi model \cite{MQRM} in which two nearly degenerate atomic manifolds couple to a single oscillator mode through a general complex coupling matrix. By working within the adiabatic regime \cite{RegimesQRM_1,Xie2017,Braak2017}, we derived an approximate spectrum 
accounting for arbitrary bright and dark decompositions determined by the singular value spectrum of the light--matter coupling matrix. Intraband detuning effects within the bright sector were incorporated perturbatively, to first order in the intraband energy spread $\varepsilon$. 
Within this framework, we obtained a closed-form expression for the thermal QFI, characterising the thermal sensitivity of our model. This naturally decomposes into contributions associated with intra-doublet bright excitations, inter-doublet bright--bright processes, and bright--dark population transfer. This allows us to pin down each spectral feature of the QFI to a specific class of transitions. By retaining QFI terms corresponding to the energy levels up to a controlled cutoff, we manage to remain accurate over the temperature range of interest while making computation, in particular, large-scale Monte Carlo sampling of complex models, computationally tractable.

We found that two limiting cases exhibit distinct thermometric mechanisms under generic random light--matter couplings with Ginibre statistics. In the dark-manifold saturation limit, a small bright sector coupled to a highly degenerate dark band generates a dominant high-temperature QFI peak governed by the interplay between a small bright sector and a large dark band. As the dark degeneracy increases at fixed bright structure, the peak QFI associated to bright--dark transitions approaches the ultimate thermometric precision \cite{Therm_3_optimal}. In particular, the associated sensitivity peak remains stable under disorder in both intra-band detunings and light--matter couplings. We additionally found that, given a frustrated dark saturation setup, with a small number of bright doublets, increasing coupling to the intermediate coupling regime can improve relative peak thermometric performance compared to an ideal thermometer. 

In the bright-manifold saturation limit, where all states are bright and many doublets contribute, thermometric sensitivity is distributed across a dense set of bright inter-doublet transition energies generated by the spread of singular values and their pairwise differences. The resulting QFI is broadband by construction and becomes increasingly more robust against variations in model parameters as the manifold size grows. 
In this regime, the peak sensitivity is typically below that achieved by the dark manifold saturation, but we showed overall that large bright manifolds can support broad temperature windows due to the enhanced density of thermally active inter doublet gaps.

Overall, our results show that multilevel light--matter coupled systems can realise two complementary thermometric advantages: we achieve high peak sensitivities via dark manifold saturation but also, may attain increasingly broad and stable response via bright manifold saturation. The analytic structure provided by the multilevel adiabatic approximation and the closed form QFI expression offers a practical route to characterising and optimising such thermometers in large, structured multilevel quantum Rabi models without the cost of full numerical diagonalisation. Future work could extend these conclusions beyond the adiabatic regime, incorporate dissipative effects, and connect the specific bright and dark saturated design principles more directly to specific cavity-QED platforms.

\section*{Acknowledgments}
We thank Sven Gnutzmann and Tommaso Tufarelli for very helpful discussions. L.A.C. acknowledges support from Ministerio de Ciencia e Innovación and
European Union (FEDER) (PID2022-138269NB-I00) and Ramón y Cajal Fellowship
(RYC2021325804-I), funded by MCIN/AEI/10.13039/501100011033 and ‘NextGenerationEU’/PRTR. J.G. and A.D.A. acknowledge support from the Leverhulme Trust under Research Project Grant Ultra-Cool Mechanics (RPG-2023-177).
G.A.~acknowledges funding from the UK Research and Innovation (UKRI EPSRC Grant No.~EP/X010929/1).

\section*{Data Availability}

The data that support the findings of this study were generated by numerical simulations. The source code used to generate the simulations is publicly available \cite{zenodorepo}.

\appendix

\section{Computing the QFI}\label{app:qficomputing}

In the strong coupling regime the lowest AA energies are not guaranteed to be those with the lowest $n$. For instance, in Fig.~\ref{fig:AAverificationSPEC} we see that bright states in the $n\geq1$ manifolds reach energies lower than the $n=0$ dark state manifold as $g/\omega_f$ grows. 
Our system Hamiltonian and, therefore, our QFI both contain an infinite number of terms due to the oscillator mode. Crucially, however, for a thermal state, contributions from higher energies become exponentially suppressed \cite{Therm_Review,quantumestimation}. For computing the QFI using Eq.~\eqref{QFI_formula} we thus adopt an energy-ordered AA truncation method, whereby we retain all terms that correspond to levels up to and including the $n=\Theta$ dark manifold, and compute $\mathcal{F}_T$ from this set. For the bright manifold, this can be accomplished by considering the 0-th order AA energies and using the condition
\begin{eqnarray}
E_k(n)\leq\Lambda_{k\emptyset}^{\Theta} \Longleftrightarrow n \leq \frac{\lambda_k^2}{\omega_f^2}+\frac{\omega_ap}{2\omega_f}+\Theta = \zeta(\Theta,k),
\end{eqnarray}
derived from Eqs.~\eqref{0thorderAAbright} and \eqref{AAdarkenergies}.
For a given $\Theta$ cutoff parameter, and $k$-th bright doublet index, we retain all $n$ integer oscillator index terms that fall below $\zeta(\Theta,k)$. Across all parameter sets tested, retaining terms corresponding to levels up to (and including) the first few dark manifolds was sufficient to guarantee convergence to the numerically exact QFI for our relevant temperature range. For the simple validation of our QFI expression, we retained energy levels in the full diagonalised model up to a much higher cutoff than in the evaluation of our formula. Increases to the cutoff $\Theta$ yield asymptotically slower improvements for higher temperatures while strong-coupling levels require progressively higher truncation and therefore, more computing power. 
For all methods it is also advisable to use arbitrary floating point precision to ensure accurate results at smaller temperatures.

Figure~\ref{fig:AAverificationQFI} validates our AA methodology for computing $\mathcal{F}_T$. In the weak-coupling case ($g=0.8\,\omega_f$ in Fig.~\ref{fig:AAverificationQFI}(a)), the AA curve reproduces the exact result at low temperatures but begins to underestimate it on the warm side, because the truncated partition function omits higher-energy manifolds whose larger excitation gaps become relevant once the temperature is increased. The leftmost peak corresponds exactly to the TLS-predicted temperature $T^*\approx0.013\,\omega_f$, set by the $n=0$ bright doublet gap $\delta^* \simeq \Lambda_{1+}^0-\Lambda^0_{1-} \approx0.054\,\omega_f$. The secondary bump at higher $T$ is likewise attributable to the energy difference between the primary doublet and the next bright--dark manifold, and is also reproduced. In both figures the intraband detuning $\varepsilon = 0.1\,\omega_a$ has no noticeable effect on the difference between the exact calculation and analytical estimate. In the strong coupling regime (Fig.~\ref{fig:AAverificationQFI}(b)), the formula reproduces the exact curve similarly across the useful temperature range. 

\section{Ideal thermometer}
\label{idealquantumtherm_appendix}
In this section we present the `ideal quantum thermometer' as introduced in \cite{Therm_3_optimal}: an equilibrium probe that maximises temperature sensitivity for a given Hilbert space dimension and target operating temperature. We use it as a benchmark to assess how closely an engineered MQRM approaches the ideal.

The probe is comprised of an atom with single nondegenerate ground state at energy 0, and a $D$-fold degenerate excited manifold at energy $E$. Introducing the dimensionless inverse temperature $x = \beta E$, the QFI for the model is then given by
\begin{eqnarray}
    \mathcal{F}_T^I = \frac{x^4 e^x}{E^2}\frac{D}{(D+e^x)^2}.
\end{eqnarray}
The QFI exhibits a single interior maximum at $x>4$, which solves the stationarity condition 
\begin{eqnarray}
    \partial_{x} \ln{(\mathcal{F}_T^I)} = 0 \Longleftrightarrow x = \ln\left(D\frac{x + 4}{x - 4}\right)\approx \ln(D).
    \label{stationarityxstarideal}
\end{eqnarray}
Using this stationarity condition, we can substitute into $\mathcal{F}_T^I$ to find the peak scaling to be approximately
\begin{eqnarray}
    \mathcal{F}_T^{*I} \approx \frac{(\ln D)^4}{4 E^2},.
    \label{idealpeakscaling}
\end{eqnarray}
showing how an optimal thermometer scales strongly with increasing Hilbert space dimension. It can also be shown that as $D$ increases, the ideal QFI becomes narrower, becoming relatively less sensitive to 'near-ideal' temperatures. In the main text we match the relevant energy gap and Hilbert space dimension to the gap of interest in the MQRM, so that any differences reflect the different model structures rather than parameter choices.

\begin{figure}[h]
    \centering
    \includegraphics[width=0.9\linewidth]{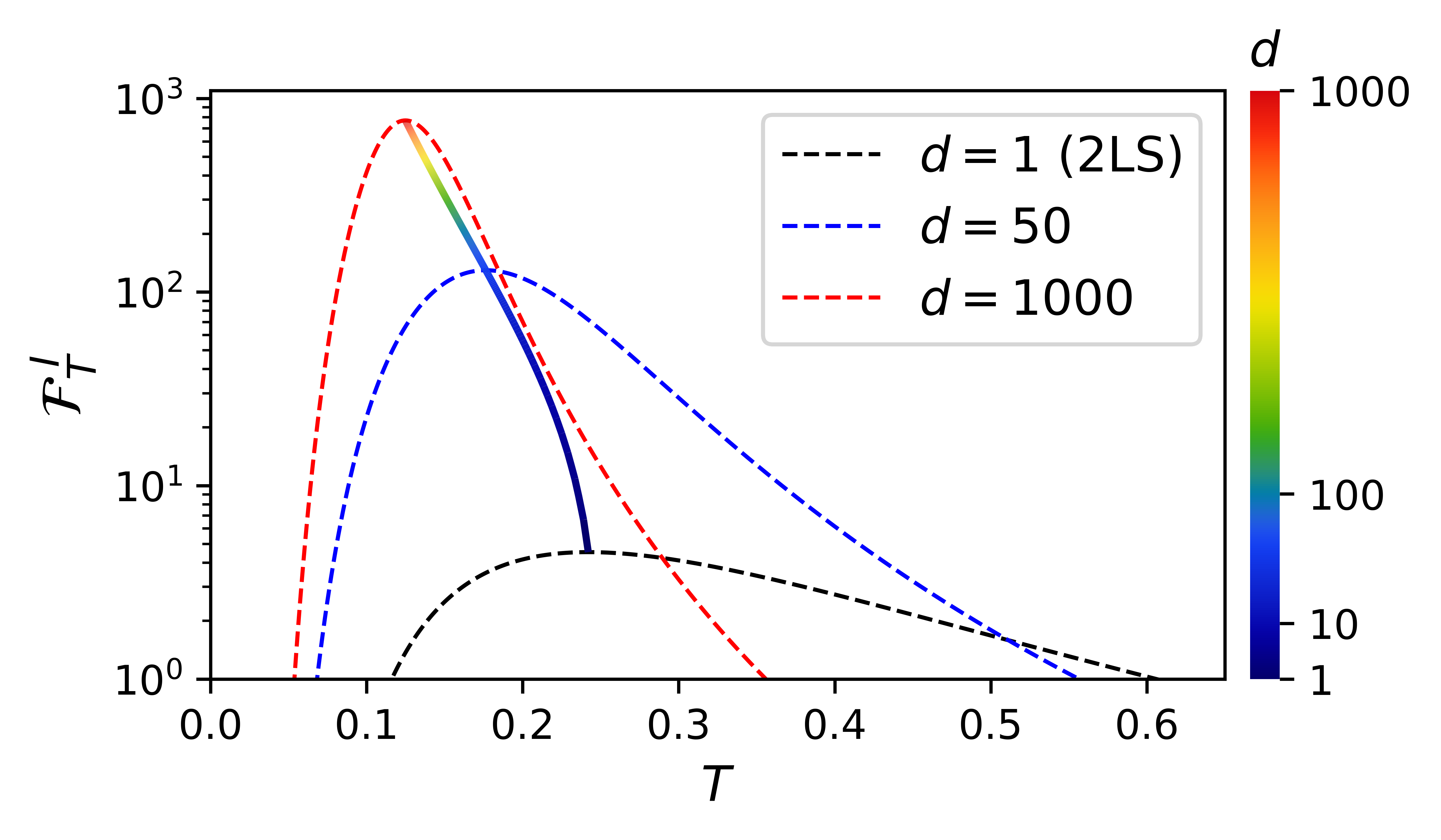}
    \caption{\textbf{Quantum Fisher information $\mathcal{F}^I_T$ versus temperature $T$ for the ideal probe at several excited-state degeneracies $d$} for a fixed energy gap $E=1$ (dashed: $d=1$ (2-level system), 50, 1000). The coloured solid curve traces the locations of maxima $x^*$ solving the stationarity condition~\ref{stationarityxstarideal}, with colour representing excited degeneracy $d$. As $d$ grows, the optimal temperature shifts downwards, peak value increases corresponding to the peak scaling equation~\ref{idealpeakscaling}, and the QFI becomes narrower and less responsive to near-ideal temperatures.}
    \label{fig:app_ideal_qfi}
\end{figure}

\section{Most likely eigenvalues of random Wishart matrices using Laguerre polynomials}
\label{laguerreappendix}
We present a self-contained derivation of the standard result that the most likely ordered set of eigenvalues of a Wishart matrix coincide with the zeros of an appropriately parameterised generalised Laguerre polynomial \cite{LagWish_1,LagWish_2,Lagwish_3}. In the main text, we use this to derive the average QFI of an ensemble of MQRM's given randomised coupling matrices and intraband detunings. 

Let $\bL\in\mathbb{C}^{n\times m}$ with $m\leq n$ (where the case $m>n$ is realised by swapping $m,n$). We define the unnormalised complex Wishart matrix $\boldsymbol{W} = \bL \bL^\dagger \in \mathbb{C}^{n\times n}$, and let $X_1>X_2>...>X_m$ denote its $m$ nonzero eigenvalues. The joint density of ordered eigenvalues for the $\beta$-ensembles is then given by \cite{LagWish_1}
\begin{equation}
    f_\beta(X) \propto \prod_{i<j}^m|X_i-X_j|^\beta\prod_i^m X_i^{\alpha\beta/2}e^{-\beta X_i/2},
\end{equation}
with $\alpha = n-m+1-2/\beta$, and $\beta$ encoding the underlying matrix symmetry. In our case $\beta=2$ corresponds to complex Wishart, but $\beta=1,4$ encode real and quaternionic entries respectively. The most likely set of eigenvalues $X^* = (X_1^*,X_2^*,...,X_m^*)$ from this density solves the stationarity conditions $\partial_{X_k} \ln f_\beta = 0$. Writing out 
\begin{equation}
    \ln{f_\beta(X)} = \beta\sum_{i<j}\ln{|X_i-X_j|} + \sum_i\frac{\alpha\beta}{2}\ln{X_i}+\sum_i \frac{\beta X_i}{2} + \text{const},
\end{equation}
we find, for each $k$ and $\beta\neq0$, 
\begin{equation}
    \partial_{X_k} \ln f_\beta= \sum_{j\neq k}\frac{1}{X_k-X_j} +\frac{\alpha}{2X_k}-\frac{1}{2}=0.
    \label{jdstat}
\end{equation}

\begin{figure}[t]
    \centering \vspace*{0.5cm}
    \includegraphics[width=1\linewidth]{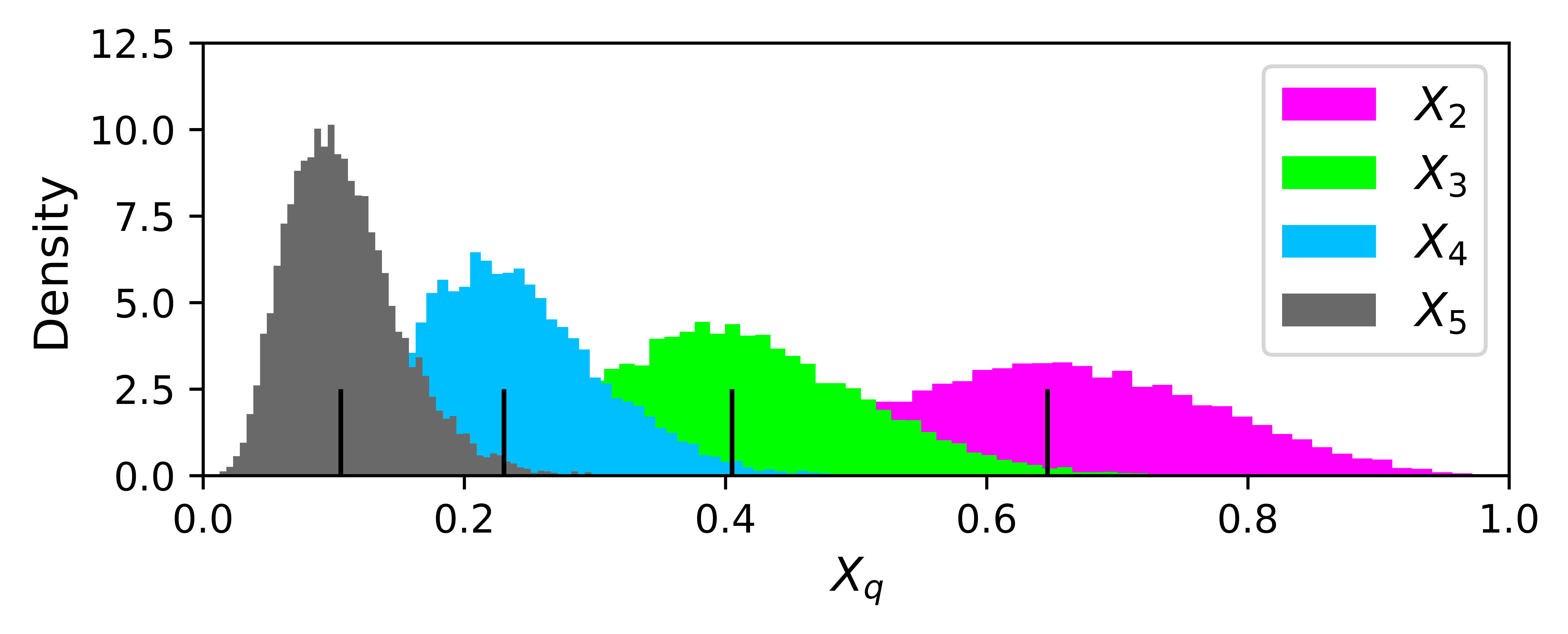}
    \caption{\textbf{Monte-Carlo validation of ordered secondary eigenvalues} $X_k$ ($k>1$) using 10000 trials for random $m=5,n=10$ independently renormalised Wishart matrices, compared to asymptotic estimates (black lines) derived using the Laguerre-polynomial approximation~\ref{laguerreeigenvaluecondition}. The plot shows that the Laguerre estimates for the mode of individual eigenvalues coincides with the peak of the respective Monte-Carlo sampled eigenvalue distributions.}
    \label{lagpolyplot}
\end{figure}

We know that the generalised Laguerre polynomial $\mathcal{L}^{\alpha'}_{m'} (x)$ is the degree-$m'$ solution of 
\begin{equation}
    xy''(x) + (\alpha'+1-x)y'(x) + m'y(x)=0,
\end{equation}
with real, simple zeros $r_1>r_2>...>r_{m'}>0$. Setting $y(x) = \prod_i^m(x-r_i)$ and using the Stieltjes relations for its respective zeros yields  
\begin{equation}
    \sum_{j\neq k}\frac{1}{r_k-r_j} + \frac{\alpha'+1}{2r_k}-\frac{1}{2}=0.
    \label{srstat}
\end{equation}
Comparing Eq.~\eqref{jdstat} and Eq.~\eqref{srstat} shows that the mode of the Wishart eigenvalue corresponds to the zeros of a generalised Laguerre polynomial with matched parameters given by 
\begin{equation}
    m=m', \quad \alpha'= \alpha-1 = n-m-\frac{2}{\beta},
\end{equation}
i.e.,
\begin{equation}
    X_k^* \text{ are the zeros of } \mathcal{L}^{n-m-2/\beta}_{m} (x) \text{ (ordered decreasing)}.
    \label{laguerreeigenvaluecondition}
\end{equation}

In the main text the nonzero singular values $\lambda_k$ of $\bL$ are related to the Wishart eigenvalues by $X_k = \lambda_k^2$. These relationships provide accurate most likely estimates for the singular spectrum. Throughout the main text we use this Laguerre-polynomial approximation to generate modal coupling profiles for ensembles corresponding to large atoms without having to resort to brute force Monte-Carlo sampling. We validate the spectra obtained using this method to Monte-Carlo densities after primary eigenvalue renormalisation $X_k/X_1$ in Fig.~\ref{lagpolyplot}.

\bibliography{bibliography}

\end{document}